\documentclass[journal]{IEEEtran}

\ifCLASSINFOpdf
\else
\usepackage[dvips]{graphicx}
\fi
\usepackage{url}

\usepackage{graphicx} 
\usepackage{subfigure}
\usepackage{graphicx,amssymb,amstext,amsmath}
\usepackage{multirow}
\usepackage{subcaption}
\usepackage[ruled,vlined]{algorithm2e}
\usepackage{bbm}
\usepackage{epstopdf}
\usepackage{amsfonts} 
\usepackage{color}
\usepackage{amssymb}
\usepackage[noadjust]{cite}
\usepackage{bm,nicefrac}
\usepackage{amstext}
\usepackage{ntheorem}
\usepackage{flushend}

\usepackage{mathtools}
\usepackage{algorithmic}
\usepackage[font=small]{caption} 

\newcommand{\bI}{{\bf I}}

\newcommand{\Real}{\mathbb{R}}
\def\x{{\mathbf x}}
\def\y{{\mathbf y}}

\def\X{{\mathbf X}}

\newcommand{\Var}{\mathrm{Var}}

\begin{document}
	
	\title{Hybrid Population Monte Carlo}
	
	\author{Ali Mousavi and  V\'ictor Elvira, \IEEEmembership{Senior Member, IEEE}
		\thanks{A. Mousavi is with Department of Computer Engineering, Neyshabur Branch, Islamic Azad University, Neyshabur, Iran (e-mail: mousavi@iau-neyshabur.ac.ir).}
		\thanks{V. Elvira is with the School of Mathematics, University of Edinburgh, United Kingdom (e-mail: victor.elvira@ed.ac.uk).}}
	
	\markboth{}
	{Shell \MakeLowercase{\textit{et al.}}: Bare Demo of IEEEtran.cls for IEEE Journals}
\maketitle

	\begin{abstract}
		%Importance sampling (IS) is a powerful Monte Carlo (MC) technique for approximating intractable distributions and integrals with respect to them, especially in the context of Bayesian inference. The performance of IS highly depends on the appropriate choice of the so-called proposal distribution where the samples are generated from. Adaptive IS (AIS) method iteratively estimates the target by adapting the proposal distribution. A family of AIS algorithms, known as population Monte Carlo (PMC), adapt a population of proposal distributions to improve the approximation of the target distribution. Recent works in the area of AIS generally focus improving the proposal adaptation procedure for high-dimensional problems. Moreover, the multi-modality of the target is in general a limitation difficult to overcome. 
		Importance sampling (IS) is a powerful Monte Carlo (MC) technique for approximating intractable integrals, for instance in Bayesian inference. The performance of IS relies heavily on the appropriate choice of the so-called proposal distribution. Adaptive IS (AIS) methods iteratively improve target estimates by adapting the proposal distribution. Recent AIS research focuses on enhancing proposal adaptation for high-dimensional problems, while addressing the challenge of multi-modal targets.
		In this paper, a new class of AIS methods is presented, utilizing a hybrid approach that incorporates weighted samples and proposal distributions to enhance performance. This approach belongs to the family of population Monte Carlo (PMC) algorithms, where a population of proposals is adapted to better approximate the target distribution. The proposed hybrid population Monte Carlo (HPMC) implements a novel two-step adaptation mechanism. In the first step, a hybrid method is used to generate the population of the preliminary proposal locations based on both weighted samples and location parameters. We use Hamiltonian Monte Carlo (HMC) to generate the preliminary proposal locations. HMC has a good exploratory behavior, especially in high dimension scenarios. In the second step, the novel cooperation algorithms are performing to find the final proposals for the next iteration.  
		HPMC achieves a significant performance improvement in high-dimensional problems when compared to the state-of-the-art algorithms. We discuss the statistical properties of HPMC and show its high performance in two challenging benchmarks.
		
	\end{abstract}
	

	\IEEEpeerreviewmaketitle

	\section{Introduction}
	\IEEEPARstart{I}{n} the field of statistical signal processing, many tasks require the computation of expectations with respect to a probability density function (PDF).
	Unfortunately, deriving analytical solutions for these expectations is often unfeasible in many real-world challenging problems. 
	A relevant example occurs in the realm of Bayesian or probabilistic inference. Here, the objective is to formulate a posterior distribution involving unknown parameters. 
	This distribution is constructed by combining data, through a likelihood model, and previous information represented by the prior distribution.
	However, a persistent challenge arises—the posterior distribution is frequently beyond reach, primarily due to the complexity of computing the marginal likelihood. 
	This complexity poses significant hurdles in numerous practical scenarios.
	Various approximation techniques are available to tackle this issue, the most popular of which is the Monte Carlo (MC) methodology \cite{robert2monte, liu2001monte, owen2013monte}. 
	MC methods have demonstrated their efficiency in addressing this problem by generating a collection of random samples. These samples serve the purpose of approximating a target probability distribution and integrals related to it.
    Importance sampling (IS) is a popular class of MC methods. It is theoretically sound, simple-to-understand, and widely applicable. 
	The basic IS involves: a) generating samples by simulating from another distribution, called proposal, (b) weighting each sample based on the discrepancy between the target and the proposal distribution, and (c) approximating integrals or density using weighted samples..
	The efficiency of the IS estimators is driven by the choice of the proposal distribution, which is a hard and very relevant problem. 
	This choice is even more challenging in situations where we only have access to the evaluation of an unnormalized version of the posterior distribution, as in the case of Bayesian inference.
	One significant methodological advancement to address the limitations of IS is multiple importance sampling (MIS), where samples are generated from multiple proposal distributions. 
	%There are several reasons for using multiple proposals. Firstly, when the target distribution is multi-modal, a set of proposals can better capture its complexity compared to a single proposal. Secondly, selecting a single optimal proposal distribution a priori is often challenging. In such cases, employing multiple proposals facilitates more efficient exploration of the parameter space. 
	A comprehensive MIS framework was recently introduced in \cite{elvira2019generalized}, demonstrating various sampling and weighting schemes that can be utilized.
	Despite the benefits of MIS, adapting the proposals can be even more advantageous, especially in high-dimensional scenarios. The intuition behind adaptive IS (AIS) methods is to adapt one or a mixture of proposals, iteratively improving the quality of the estimator by better fitting the proposals \cite{oh1992adaptive}.
	%
	%The research in AIS continues being very active and many crucial challenges remain open (see \cite{bugallo2017adaptive,elvira2021advances} for recent surveys).
	
	AIS methods gained prominence in MC computations following the introduction of the population Monte Carlo (PMC) \cite{cappe2004population} family of algorithms. PMC uses a resampling step based on previous weighted samples to adapt a population of proposal PDFs \cite{douc2005comparison,li2015resampling}. Several AIS algorithms have been proposed since the publication of PMC. We refer the reader to \cite{bugallo2017adaptive,elvira2021advances} for an exhaustive review.
	AIS methods are diverse and widely used, with key differences in how proposals are selected, weighted, and adapted. A critical step is proposal adaptation, which can be categorized into three main approaches. The first category adapts proposals using the latest samples, as in PMC methods \cite{cappe2004population}, with enhancements such as N-PMC \cite{koblents2015population}, M-PMC \cite{cappe2008adaptive}, and DM-PMC \cite{elvira2017improving} and more advanced methods based on local curvature of the target like \cite{elvira2019langevin,elvira2022optimized}. The second category utilizes all past samples, as in AMIS \cite{cornuet2012adaptive} and APIS \cite{martino2015adaptive}, or sequential Monte Carlo methods like those in \cite{nguyen2015efficient,karamanis2024persistent}. The third category employs hierarchical adaptations, such as GAPIS \cite{elvira2015gradient}, ULA-based methods \cite{schuster2015gradient, fasiolo2018langevin}, and advanced layered AIS methods like LAIS \cite{martino2017layered}, MCIS \cite{klebanov2020markov}, and HAIS \cite{mousavi2021hamiltonian}. The proposed method in this paper belongs to the PMC family, leveraging a hybrid adaptation approach from both the first and third categories.
	
	One challenging issue in layered AIS algorithms is the lack of sufficient cooperation during the adaptation procedure to generate proposal parameters for the next iteration. Additionally, recent layered methods often require an external proposal PDF that is unrelated to the target PDF. For instance, the PI-MAIS and $ I^2 $-MAIS algorithms in LAIS use an independent traditional Metropolis-Hastings (MH) algorithm for adaptation. 
	%The random-walk behavior of many traditional MCMC algorithms like MH makes Markov chain convergences to the target distribution inefficiently, especially in high-dimensional multi-modal distributions. As a result, one main drawback of LAIS is lack of exploration capability which is due to independent MH algorithm for adaptation. 
	HAIS, as a notable extension of LAIS, introduces a two-layered AIS that retains advantageous features from both Hamiltonian Monte Carlo (HMC) \cite{duane1987hybrid, neal2011mcmc} and AIS. HAIS employs a more efficient adaptive procedure, offering additional benefits such as parallelization and cooperation via resampling for efficient inference.
	A major limitation of the adaptation procedures in layered methods is their exclusive dependence on samples or proposal parameters generated within the current iteration. Specifically, the adaptation and sampling steps in these AIS methods are decoupled, meaning the samples generated for estimation are not utilized in the adaptation process. Another significant challenge in layered AIS algorithms is the lack of sufficient cooperation during the adaptation phase to generate proposal parameters for subsequent iterations. Additionally, recent layered methods often require an external proposal PDF that is unrelated to the target PDF.
	
	{In this paper, we present hybrid-PMC (HPMC), a new class of AIS methods that employs a hybrid adaptation strategy, leveraging both weighted samples generated from the proposals and the proposal parameters (see Fig. \ref{figure_adaptive_families}-d). 
	This method extends the PMC framework by enhancing the adaptation step to achieve an effective balance between local and global exploration, making it particularly suitable for high-dimensional, multi-modal targets. Moreover, it retains the advantageous exploratory behavior of HMC, especially in high-dimensional scenarios.
	The proposed algorithm introduces a two-step adaptation mechanism. In the first step, preliminary parameters (locations) from the proposals are generated based on two sets: {\it i) weighted samples generated from the proposals}: These enrich the diversity in the population of proposal locations, increasing the likelihood of discovering local features of the target; {\it ii) location parameter of the proposals}: These guide the proposals toward unexplored regions, promoting global exploration. By leveraging both weighted samples and location parameters, our approach reduces the number of iterations needed to reach the desired estimate, leading to faster convergence.
		
	Another significant contribution of this work is the novel cooperation step, where information among preliminary locations is exchanged to redirect the lost proposals toward the optimal path based on the target PDF. Two novel methods are proposed for determining the final set of proposal locations for the next iteration of the AIS procedure: cooperation by resampling and cooperation by weighted mixture modeling. This cooperation step introduces a theoretically sound and consistent procedure to enhance global exploration. Notably, this cooperation step does not require any free parameter tuning. Finally, HPMC implements the DM weighting technique for both target estimation and adaptation, which has demonstrated superior performance compared to other methods in the AIS literature.
	
	In a nutshell, the contributions of this paper are as follows:
	\begin{itemize}
		\item 	We propose a new class of AIS methods that, for the first time, combines weighted samples and location parameters to achieve an effective balance between local and global exploration, significantly accelerating estimator convergence.
		\item  A novel cooperation step is introduced to redirect lost proposals, enhancing global exploration and optimizing the allocation of the computational complexity without requiring any parameter tuning.
		\item  HPMC incorporates the DM weighting technique, which demonstrates superior performance in both target estimation and proposal adaptation compared to existing techniques
	\end{itemize}
     }
	The rest of the paper is organized as follows. Section \ref{sec_problem}, describes the background in Monte Carlo. The novel HMPC method is presented in Sections \ref{sec_HAIS}-A to C, and its theoretical foundation is discussed in Section \ref{sec_Discussion}. The paper concludes with three numerical examples and a discussion on computational complexity comparison in Section \ref{sec_Experiments}. Finally, some concluding remarks is explained in Section \ref{sec_Conclusion}.

	\section{Background in Monte Carlo}
	\label{sec_problem}
	%\subsection{\cred{Bayesian inference}}
	In the Bayesian framework, the variable of interest $ \x\in \mathbb{R}^{d_x} $ with related measurements $ \y \in \mathbb{R}^{d_y} $, is characterized through the posterior probability function or the target probability density function (PDF) known as the 
	\begin{equation}
		\widetilde{\pi}(\x|\y)=\frac{\ell(\y|\x)p_0(\x)}{Z(\y)}\propto \pi(\x)=\ell(\y|\x)p_0(\x),
	\end{equation}
	where $ \ell(\y|\x) $ is the likelihood function, $ p_0(\x) $, represents the prior PDF, and $ Z(\y) $ is the model evidence. We alleviate the notation by dropping $\y$, since our goal is to address more general frameworks beyond the Bayesian inference. 

	In many applications the goal is to compute the integral
	\begin{equation}
		\begin{aligned}
					I =\int_{\mathbb{R}^{d_{x}}} f(\x)\widetilde{\pi}(\x)d\x=\frac{1}{Z}\int_{\mathbb{R}^{d_{x}}} f(\x){\pi}(\x)d\x,
		\label{integralI}
		\end{aligned}
	\end{equation}
	where $ f: \mathbb{R}\rightarrow \mathbb{R}^{d_x} $ is any integrable function w.r.t. $\widetilde{\pi}(\x)$ and $Z=\int \pi(\x)d\x$. In many cases of interest, it is impossible or too complicated to find an analytical solution to this integral, in which case we resort to the Monte Carlo (MC) methods. 
	
	\subsection{Importance sampling}
	Importance sampling (IS) is a MC technique. To comprehend IS, we start by rewriting Eq. \eqref{integralI} as 
	\begin{equation}
		I=\mathbb{E}_{\widetilde{\pi}(\x)}\left[f(\X)\right]=\mathbb{E}_{q(\x)}\left[f(\X)\frac{\widetilde{\pi}(\X)}{q(\X)}\right],
	\end{equation}
%	\begin{equation}
%		\begin{aligned}
%			I = \int_{\mathbb{R}^{d_{x}}} f(\x)\widetilde{\pi}(\x)d\x
%			= &\int_{\mathbb{R}^{d_{x}}} \frac{\widetilde{\pi}(\x)}{q(\x)}f(\x)q(\x)d\x\\
%			=&E_{q(\x)}\left[\frac{\widetilde{\pi}(\X)}{q(\X)}f(\X)\right]
%		\end{aligned}		
%	\end{equation}
	%
	where $\mathbb{E}(.)$ denotes expectation for $\X \sim q(\x)$ and $q(\x)$ is the so-called proposal distribution with the constraint that $q(\x)>0$ for all $\x$ where $\widetilde{\pi}(\x)f(\x)\neq 0 $. The basic idea of IS is to approximate this integral by generating $M$ random samples $\x_m$ from a simple proposal PDF $q(\x)$, and each sample is associated with an importance weight $w_m$. The IS procedure involves two fundamental steps:
	\subsubsection{\textbf{Sampling}}
	$M$ iid samples are generated as 
	\begin{equation}
		\x _m\sim q(\x), \; m=1, \cdots,     M. \nonumber
	\end{equation}
	\subsubsection{\textbf{Weighting}}
	each sample $\x_m$ is assigned an importance weight as
	\begin{equation}
		w_m=\frac{\pi(\x_m)}{q(\x_m)}. \nonumber
	\end{equation}
	The resulting pairs of $\left\lbrace\x_m,w_m\right\rbrace^{M}_{m=1}$ are used to approximate the integral $I$ w.r.t. to the target distribution $\widetilde\pi(\x)$ as
	\begin{equation}
		\widehat{I}_{\text{IS}}(\x)=\frac{1}{MZ}\sum_{m=1}^{M}w_mf(\x_m).
		\label{eq_uis}
	\end{equation}
	Thus we can construct the \emph{unnormalized IS} (UIS) estimator $\widehat{I}_{\text{IS}}$, which is an unbiased and consistent estimator of $I$.
	When $Z$ is not available, the \emph{self-normalized IS} (SNIS) estimator can be constructed by substituting the unbiased estimator $\widehat{Z}=\frac{1}{M}\sum_{m=1}^{M}w_m$ for $Z$ in Eq. \eqref{eq_uis}, i.e., 
	\begin{equation}
		\widetilde{I}_{\text{IS}}(\x)=\frac{1}{M\widehat{Z}}\sum_{m=1}^{M}w_mf(\x_m)=\sum_{m=1}^{M}\bar{w}_mf(\x_m) .
	\end{equation}
	%where $ \bar{w}_m={w_m}/{\sum_{i=1}^{M}w_i}, m=1,\cdots,M $ are normalized importance weights (see more details in \cite{elvira2019generalized}).
	%
	SNIS  is a consistent but biased estimator and as $M$ increases, with the bias decreasing more rapidly than the variance as M increases. 
	\subsection{Multiple importance sampling}
	\label{sec_MIS}
	The IS method has been extended to scenarios where samples are drawn from multiple proposals, ${q_n(\x)}_{n=1}^N$, rather than a single one. This approach, called multiple importance sampling (MIS), introduced by Veach and Guibas \cite{veach1995optimally}, is particularly effective for fitting multi-modal target distributions.
	In standard MIS, we generate $K$ samples from each proposal, i.e.,
		\begin{equation}
		\x_{n,k}\sim q_n(\x), \; n=1,...,N , \: \textit{and} \: k=1,...,K.
		\nonumber
	\end{equation}
	Then, the generalized MIS estimator as in \cite{veach1995optimally} is given by 
		\begin{equation}
		\widetilde{I}_{\text{MIS}}(\x)=\sum_{n=1}^{N}\frac{1}{K}\sum_{k=1}^{K}g_n(\x_{n,k})\frac{f(\x_{n,k})\pi(\x_{n,k})}{q_n(\x_{n,k})},
	\end{equation}
	where $ g_n(\x)\geq 0$ is a weighting function corresponding to the $n$-th proposal which satisfies $\sum_{n=1}^{N}g_n(\x)=1$ for all $\x$.
	The most common function for $g_n(\x)$ is the \textit{balance heuristic} with
	\begin{equation}
		\begin{aligned}
		g_n(\x)&=\frac{Kq_n(\x)}{\sum_{j=1}^{N}Kq_j(\x)} \\
		&=\frac{1}{N}\frac{q_n(\x)}{\sum_{j=1}^{N}q_j(\x)}.
		\end{aligned}
	\end{equation}
	In this case, the MIS estimator $ \widetilde{I}_{\text{MIS}}(\x)$ becomes the balance heuristic estimator, i.e.,
	\begin{equation}
		\begin{aligned}
				\widetilde{I}_{\text{MIS}}(\x)&=\frac{1}{ZN}\sum_{n=1}^{N}\frac{1}{K}\sum_{k=1}^{K}f(\x_{n,k})w_{n,k},
		\end{aligned}
		\label{eq_mis}
	\end{equation}
	with $w_{n,k}={\pi(\x_{n,k})}/{\frac{1}{N}\sum_{j=1}^{N}q_j(\x_{n,k})}$. We can also refer to  Eq. \eqref{eq_mis} as the deterministic mixture MIS (DM-MIS) estimator \cite{owen2000safe}. In the standard MIS weighting scheme, we define
	\begin{equation}
		w_{n,k}=\frac{\pi(\x_{n,k})}{q_n(\x_{n,k})}, \; n=1,...,N , \: \textit{and} \: k=1,...,K,
	\end{equation}
	which corresponds to the standard MIS (s-MIS) estimator \cite{cappe2004population}. It is possible to construct a MIS estimator using either of the weighting functions. However, the MIS estimator with DM weights outperforms the estimator with standard weights. Specifically, it has been shown that $\Var[\widetilde{I}_{\text{DM-MIS}}(\x)] \leq \Var[\widetilde{I}_{\text{s-MIS}}(\x)]$ \cite{elvira2019generalized}. 
	Note that both weighting schemes involve the same number of target evaluations; however, the DM-MIS estimator requires more computational effort in terms of proposal evaluations (with $N^2$ evaluations needed for the DM weights compared to $N$ evaluations for the standard weights). Several efficient methods have been proposed in the literature to reduce computational complexity \cite{elvira2015efficient,elvira2016heretical}.
	There are various weighting functions in the literature. For further details on these weighting schemes, which lead to consistent estimators, we refer the reader to \cite{elvira2019generalized}.
	\subsection{Adaptive importance sampling}
	Adaptive importance sampling (AIS) iteratively adapts one or several proposals to approximate the target PDF. AIS methods typically involve three key steps:
	\subsubsection{Sampling}
	Generate a set of samples from one or multiple proposal PDF(s).
	\subsubsection{Weighting}
	Use a weighting function to compute the weight of each sample obtained in the previous step.
	\subsubsection{Adaptation}
	Update the proposal(s) to improve the sampling for the next iteration.

	%There are several ways to combine these $KNT$ weighted samples. A typical approach is to compute the normalized weights by considering all the weights, i.e.,
%	\begin{equation}
%		w_{n,k}^{(t)}=\frac{w_{n,k}^{(t)}}{\sum_{i=1}^{T}\sum_{j=1}^{K}\sum_{m=1}^{K}w_{j,m}^{(i)}}
%	\end{equation}
	A typical self-normalized AIS estimator to approximate the integral $I$ w.r.t. the target PDF $\widetilde{\pi}$ up to iteration $t$ is given by
	\begin{equation}
		\widetilde{I}_{\text{AIS}}(\x)=\frac{1}{KNt\widehat{Z}}\sum_{\tau=1}^{t}\sum_{n=1}^{N}\sum_{k=1}^{K}w_{n,k}^{(\tau)}f(\x_{n,k}^{(\tau)}),
	\end{equation}
	and
	\begin{equation}
		\widehat{Z}=\frac{1}{KNt}\sum_{\tau=1}^{t}\sum_{n=1}^{N}\sum_{k=1}^{K}w_{n,k}^{(\tau)},
	\end{equation}
	where $\x_{n,k}^{(\tau)}$ with $w_{n,k}^{(\tau)}$ is the $k$-th weighted sample generated from $n$-th proposal PDF at iteration $\tau$ of the AIS algorithm. 
		
\begin{figure}
		\centering
		\subfigure[The proposal parameters are adapted using the last set of drawn samples.]{
			\includegraphics[scale=0.15]{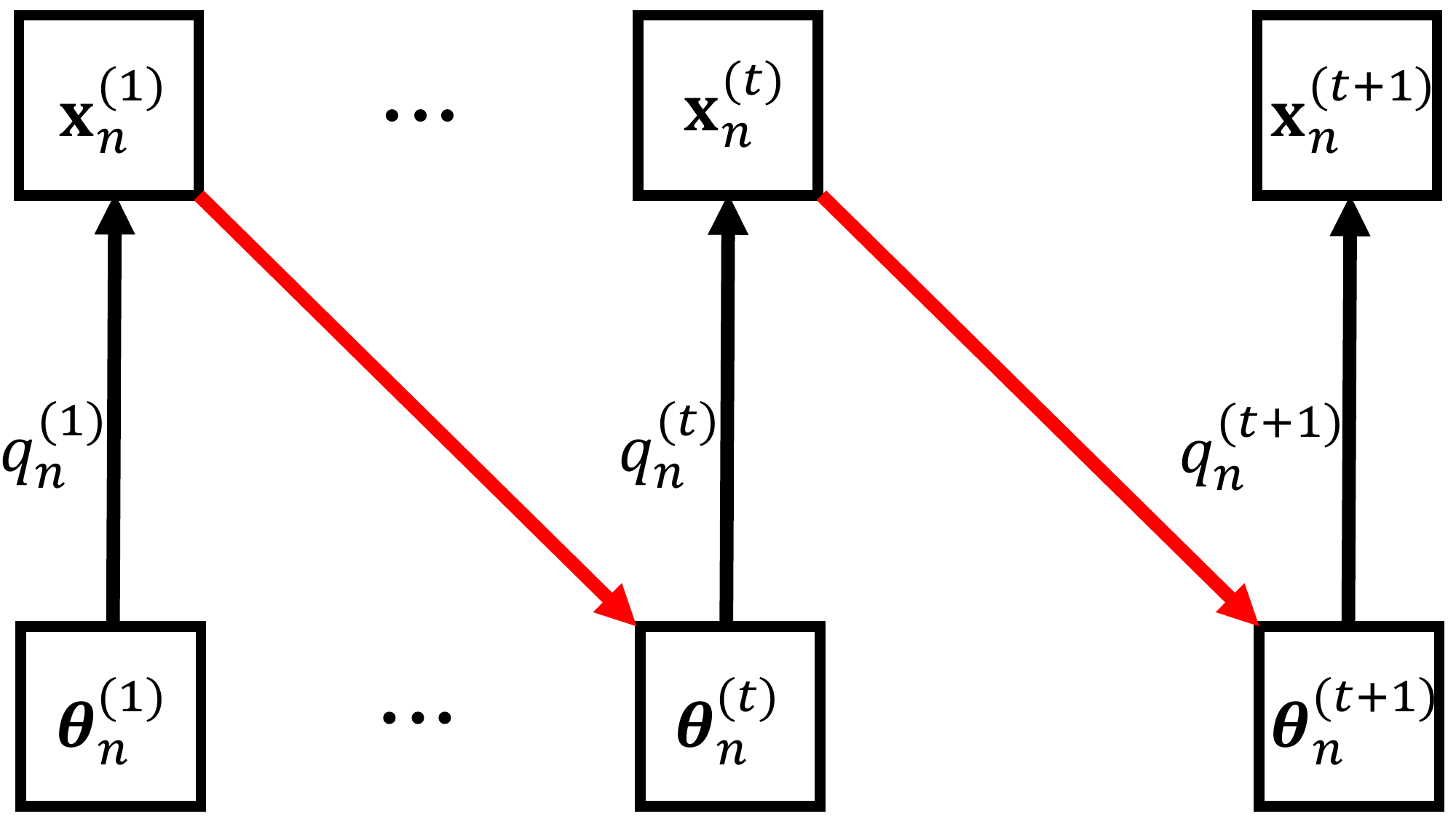}
		} \quad
		\subfigure[The proposal parameters are adapted using all drawn samples up to the latest iteration.]{
			\includegraphics[scale=0.15]{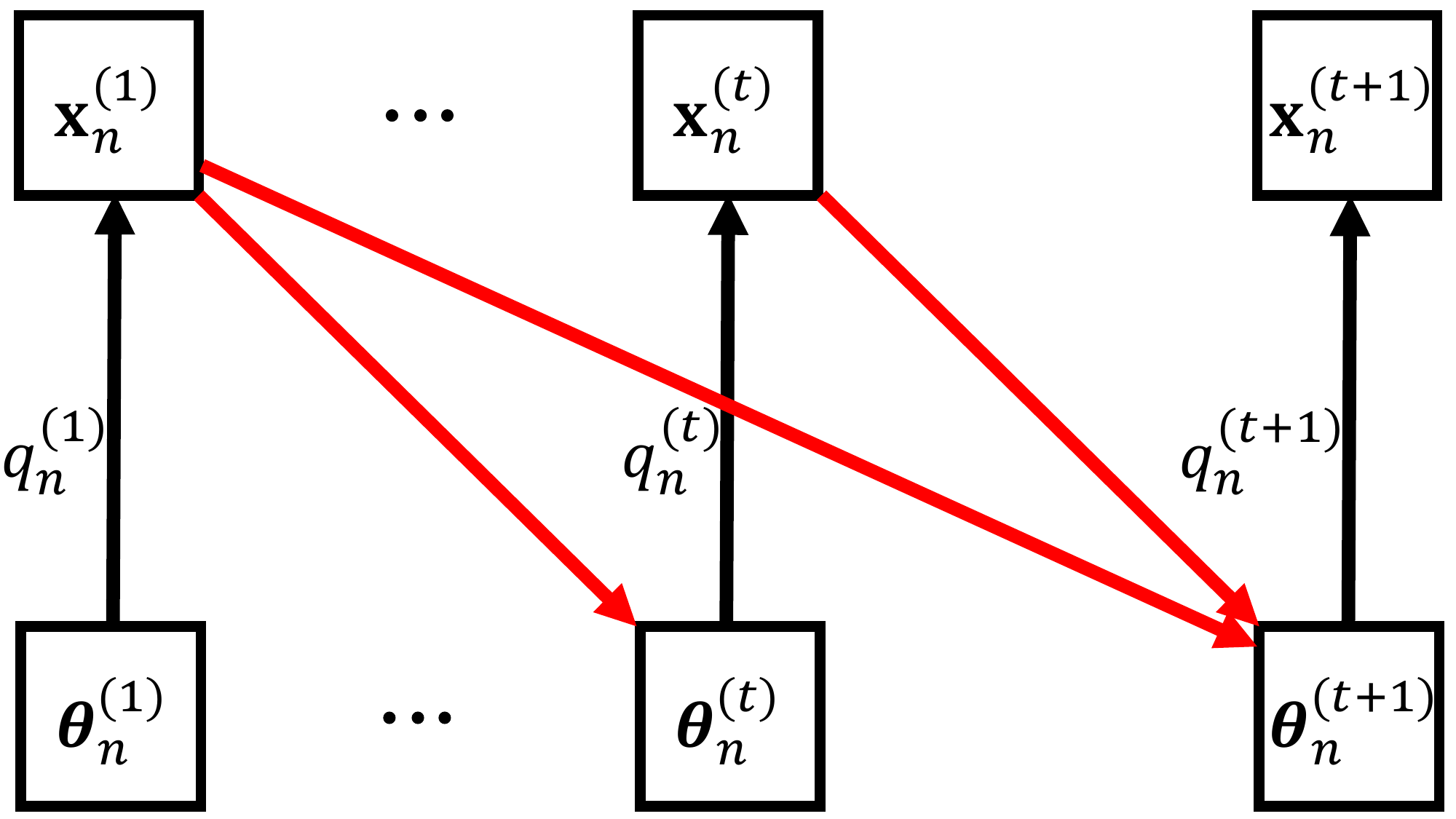}
		} \quad
		%\centering
		\subfigure[The proposal parameters are adapted using an independent process from the samples.]{
			\includegraphics[scale=0.15]{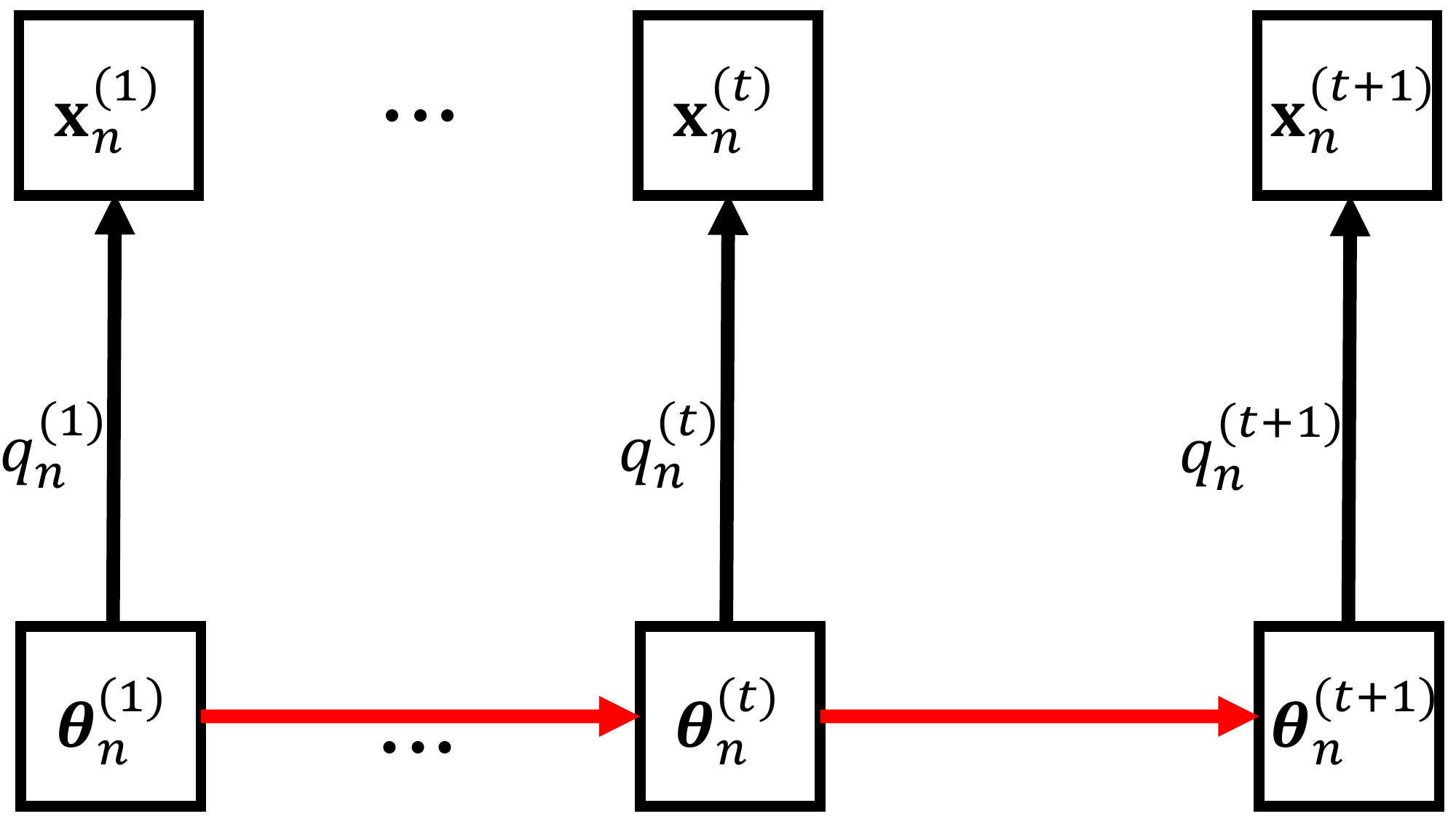}
		}
		\centering
		\subfigure[The proposal parameters are adapted using both the generated weighted samples and the location parameters of the proposals.]{
			\includegraphics[scale=0.15]{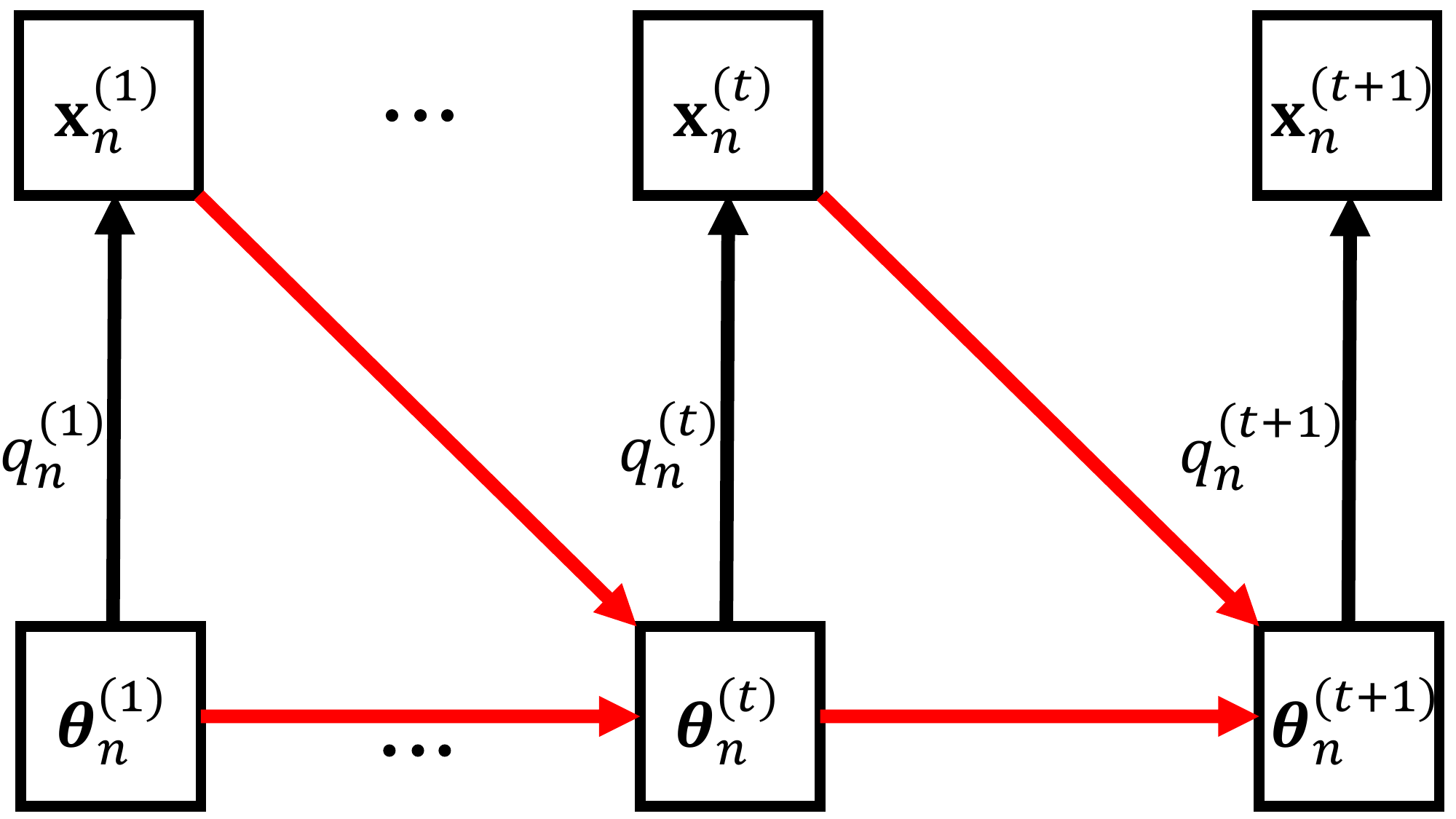}
		} \quad
		\caption{Graphical description of the three main existing classes of AIS algorithms (a-c), and the new class (d). }
		\label{figure_adaptive_families}
	\end{figure}

	AIS methods vary in proposal types, weighting schemes, and adaptation strategies, with proposal adaptation being crucial. The most significant adaptation procedures in AIS families can be classified into three distinct categories. Fig. \ref{figure_adaptive_families}(a-c) graphically illustrates the three categories, with each subplot depicting the dependency of proposal parameter adaptation and the sample generation process. The index $t$ represents the iteration number, $n$ denotes the proposal number, and $\theta_n^{(t)}$ are the parameters of the proposals to be adapted. For simplicity of presentation, we assume that only one sample is drawn from the $n$-th proposal (i.e., $K=1$). Each red arrow specifies what is used to adapt a proposal.
	The descriptions and algorithms from the literature for each category are as follows:

	(1) The proposal parameters are adapted using the last set of drawn samples (Fig. \ref{figure_adaptive_families}-a). The population Monte Carlo (PMC) methods, as introduced in \cite{cappe2004population}, belong to this category. A key characteristic of PMC methods is their use of resampling techniques to adapt the location parameters of the proposals \cite{li2015resampling}. Other approaches, such as N-PMC \cite{koblents2015population} with nonlinear weighting and M-PMC \cite{cappe2008adaptive} employing stochastic expectation-maximization, have enhanced the performance of the standard PMC method. Further advancements in weighting and resampling techniques are presented in \cite{elvira2017improving,elvira2017population}, which introduces the DM-PMC, GR-PMC, LR-PMC, and Diverse-PMC methods. Some methods \cite{elvira2019langevin,elvira2022optimized} have further improved PMC samplers for high-dimensional, multimodal target distributions by incorporating the target's local curvature after resampling. However, these strategies are inefficient in high dimensions as they require inverting the Hessian matrix of the target distribution.
	
	(2) The proposal parameters are adapted using all drawn samples up to the latest iteration (Fig. \ref{figure_adaptive_families}-b). Examples from this category include AMIS \cite{cornuet2012adaptive}, which introduces a temporal estimator based on moment matching, and APIS \cite{martino2015adaptive}, which utilizes multiple proposals across epochs. Two other works \cite{nguyen2015efficient,karamanis2024persistent} on sequential MC samplers (SMC) also construct a more diverse and rich sample set of the target posterior by leveraging samples from all iterations. 
	
	(3)  Some recent methods aim at adapting the proposals through an independent process from the past samples (Fig. \ref{figure_adaptive_families}-c). This class of algorithms could be called layered or hierarchical methods for the adaptation in IS. In the upper layer, proposal parameters are adapted while in the lower layer weighted samples are generated to estimate the target. As an example, GAPIS \cite{elvira2015gradient} exploits the gradient and the Hessian of the target in the upper layer to adapt the location and covariance parameters, respectively. Other methods, such as \cite{schuster2015gradient, fasiolo2018langevin}, adapt the location parameters by performing several steps of the unadjusted Langevin algorithm (ULA) at each sample. More advanced structures, such as LAIS \cite{martino2017layered,martino2017anti}, MCIS \cite{klebanov2020markov}, and HAIS \cite{mousavi2021hamiltonian}, have incorporated Markov chain Monte Carlo (MCMC) mechanisms for the adaptation of proposals. The method we propose in this paper belongs to the PMC framework family, utilizing a hybrid adaptation from both the first and third categories (Fig. \ref{figure_adaptive_families}-d).

	\section{Hybrid population Monte Carlo}
	\label{sec_HAIS}
	\subsection{Summary of the algorithm}
	In this section, we present the novel \emph{hybrid population Monte Carlo} (HPMC) method. HPMC is an adaptive importance sampling (AIS) method that uses multiple proposals to generate samples at each iteration, that belongs to the family of PMC algorithms. The algorithm is described in Alg. \ref{alg_HPMC}. The procedure starts by initializing $N$ (parametric) proposals, $ q_n(\x;\bm{\mu}_n^{(t)},\bm{\nu}_n) $ where $ \bm{\mu}_n^{(1)} \in \Real^{d_x}$ is the initial location parameter (e.g., the mean vector in a Gaussian PDF) and $\bm{\nu}_n$ contains the other parameters (e.g., $\bm{\nu}_n =\bm{\Sigma}_n$ is the covariance matrix in a Gaussian distribution). The algorithm runs for $T$ iterations, involving three main steps: sampling (Step 1), weighting (Step 2), and adaptation (Step 3). In Step 1 (sampling), a set of multiple proposals is used to draw samples. In Step 2 (weighting), the normalized IS weight is calculated for each sample. Finally, in Step 3 (adaptation), the parameters of all proposals are updated for the next iteration.	The adaptation process is a hybrid approach that utilizes all weighted samples generated from the proposals as well as the location parameter of the proposals. The adaptation step follows of a two-step procedure. In Step 1(a), we generate a population of preliminary locations of the proposals based on: i) weighted samples generated from the proposals and ii) location parameters. In the subsequent cooperation step (Step 2(b)), information from all preliminary locations is exchanged to generate new final proposal locations for the next iteration. Finally, the algorithm outputs the set of weighted samples to estimate $\widetilde{I}$ and $\widehat{Z}$. In the following, we discuss the steps in detail.
			\begin{algorithm}
		
		\scriptsize
		\SetAlgoLined
		\textbf{Input}: initial location, $ \{\bm{\mu}_n^{(1)}\}_{n=1}^N $, and scale, $ \{\bm{\nu}_n\}_{n=1}^N $, of proposals, $ N $ proposals, $ K $ sample per proposal, $ T $ iterations \\
		\For{$ t=1,...,T $}{
			\hspace{2pt} \textbf{Step 1 (sampling)}: draw K samples per individual proposal or mixand, 
			\begin{equation}
				\x_{n,k}^{(t)} \sim q_n(\x;\bm{\mu}_n^{(t)},\bm{\nu}_n), \; n=1,...,N , \: \textit{and} \: k=1,...,K.
				\label{Samplingg}
			\end{equation}
			\textbf{Step 2: (Weighting)}: compute the deterministic mixture (DM) weight of each sample by
			\begin{equation}
				\label{DM_weighting}
				w_{n,k}^{(t)}=\frac{\pi(\x_{n,k}^{(t)})}{\sum_{i=1}^{N}q_i(\x_{i,k}^{(t)};\bm{\mu}_i^{(t)},\bm{\nu}_i)},
			\end{equation}
			and normalize through
			\begin{equation}
				\label{DM_weighting_norm}
				\bar{w}_{n,k}^{(t)}=\frac{w_{n,k}^{(t)}}{\sum_{i=1}^{N}\sum_{j=1}^{K}w_{i,j}^{(t)}}.
			\end{equation}
			\textbf{Step 3: (adaptation)}: update location of proposals for the next iteration:
			\\ 
			\hspace{10pt}3a) \textbf{Preliminary locations generation}: obtain new population of proposal locations %$\{{\bm{\mu}}_1^{(t+1)*},{\bm{\mu}}_2^{(t+1)*},...,{\bm{\mu}}_{(M+N)}^{(t+1)*}\}$, 
			by two separate operations:\\
			\hspace{15pt} i) generate $N$ preliminary locations, $P$, by applying $N$ parallel LR sampling with the weights from Eq. (\ref{normalized_LR}). %See Section \ref{sec_HAIS}-B (Step 1-a)\\
			\\
			\hspace{15pt} ii) generate $N$ preliminary locations, $Q$, by applying one iteration of $ N $ parallel HMC blocks.  %See Section \ref{sec_HAIS}-B (Step 1-b)
			
			\hspace{10pt}3b) \textbf{Final locations generation through cooperation}: generate final population of locations, $ \{{\bm{\mu}}_1^{(t+1)},{\bm{\mu}}_2^{(t+1)},...,{\bm{\mu}}_N^{(t+1)}\} $, for the next iteration according to the methods in Section \ref{sec_HAIS}-C (Step 3(b)) 
			%by either method of weighted mixture model (Alg. \ref{alg_Coop1}) or the resampling method.   
		}
		\textbf{Output}: a set of $KNT$ samples with associated weights, $ \{ \x_{n,k}^{(t)},w_{n,k}^{(t)} \}$ to obtain the estimation.
		\caption{Hybrid population Monte Carlo}
		\label{alg_HPMC}
	\end{algorithm}
	
	\subsection{Sampling and weighting}
\begin{figure*}
	 	\centering
	 	%\begin{subfigure}[t]{0.37\linewidth}
	 	\includegraphics[width=0.9\textwidth]{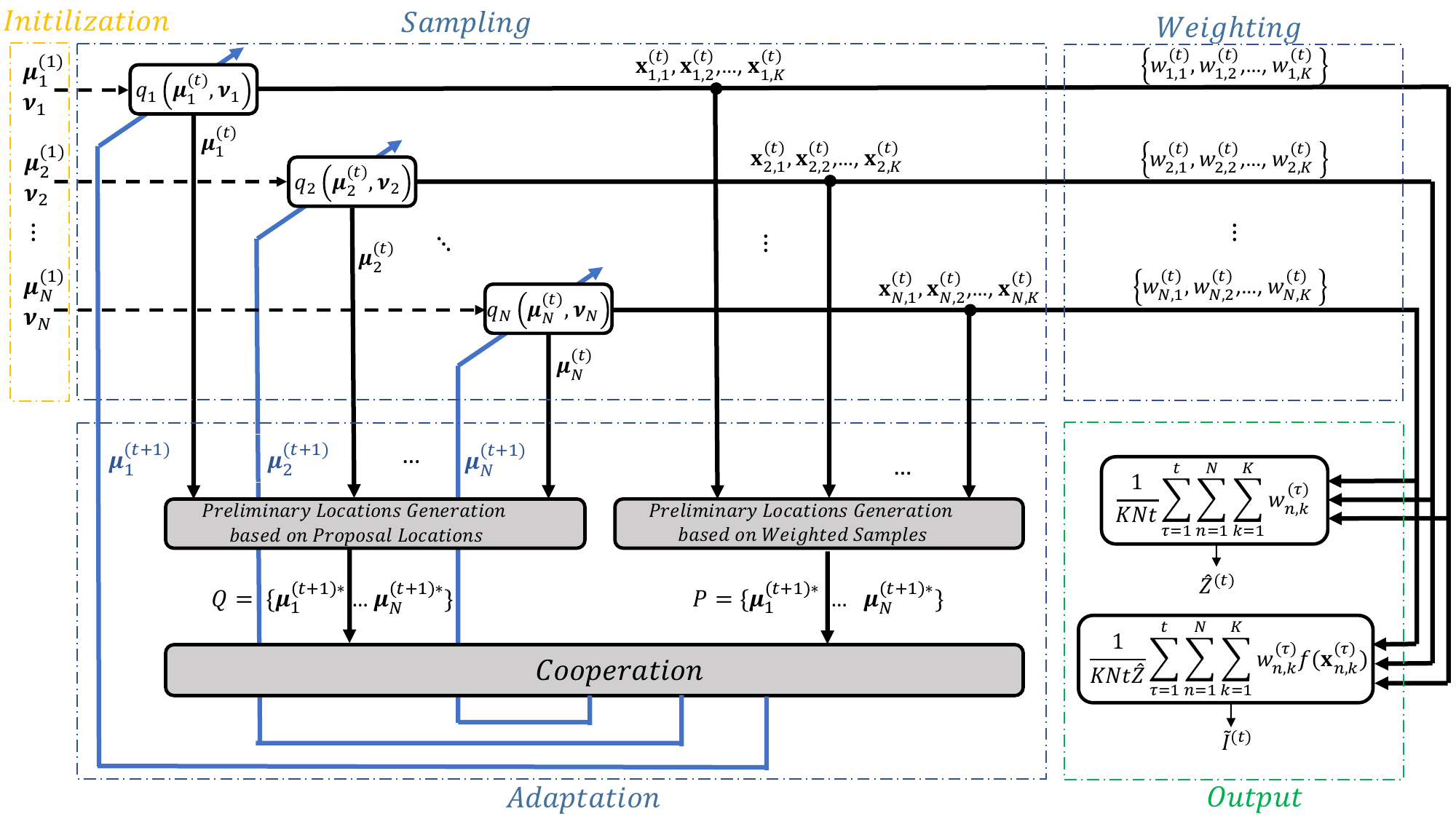}
	 	%\label{fig:flowchart}       % Give a unique label
	 	\caption{The flowchart of the proposed HPMC method illustrates the three main operations at iteration $t$, which lead to the estimated values. The process consists of three primary modules—sampling, weighting, and adaptation—along with initialization and output modules.}
	 	\label{fig:main_flowchart}
	 \end{figure*}

	 In the sampling step, at each iteration $t$, exactly $K$ samples are simulated from each of the $N$ proposals, resulting in a total of $KN$ samples.
	 The weighting operation is applied to compute the importance weights using the deterministic mixture (DM) scheme according to Eq. \eqref{DM_weighting} and then normalized through Eq. \eqref{DM_weighting_norm}. DM weighting has shown to provide an estimator with lower variance compared to the others (as discussed in Section \ref{sec_MIS}). Each of these operations is represented as a separate module in Fig. \ref{fig:main_flowchart}, illustrating the step-by-step structure of the algorithm. 
	 
	\subsection{Adaptation process}
	The proposed adaptation process is a hybrid approach that includes two steps: Step 3(a), preliminary location generation, and Step 3(b), final location generation through cooperation. Below, we provide a detailed description of this novel approach:
	
	\textbf{Step 3(a): Preliminary locations generation}
	
	We introduce two distinct operations to find two sets of preliminary locations of the proposal PDFs, $\{\bm{\mu}_1^{(t+1)*},\bm{\mu}_2^{(t+1)*},\cdots,\bm{\mu}_N^{(t+1)*}\}$, discussed as follows.

	{\it i) Preliminary locations generation based on weighted samples}
	
	We propose using the current set of $NK$ samples from Eq. (\ref{Samplingg}), weighted by Eq. (\ref{DM_weighting}), to generate the population of $N$ preliminary locations for the next iteration (as illustrated in the top-right block of the adaptation module in Fig. \ref{fig:main_flowchart}). Next, $N$ local resampling (LR) \cite{elvira2017improving} procedures are independently performed in parallel within each subset of $K$ samples. The $n$-th preliminary location $\bm{\mu}_n^{(t+1)*}$, for $n=1,\cdots,N$, is resampled from the set $\{\x_{n,k}^{(t)} \}_{k=1}^K$ using the following normalized weight as the multinomial probability mass function:
	\begin{equation}
		\label{normalized_LR}
		\bar{w}_{n,k}^{(t)}=\frac{w_{n,k}^{(t)}}{\sum_{i=1}^{K}w_{n,i}^{(t)}}.
	\end{equation}
	
\begin{figure}[h]
		\centering
		%\begin{subfigure}[t]{0.37\linewidth}
		\includegraphics[width=0.5\textwidth]{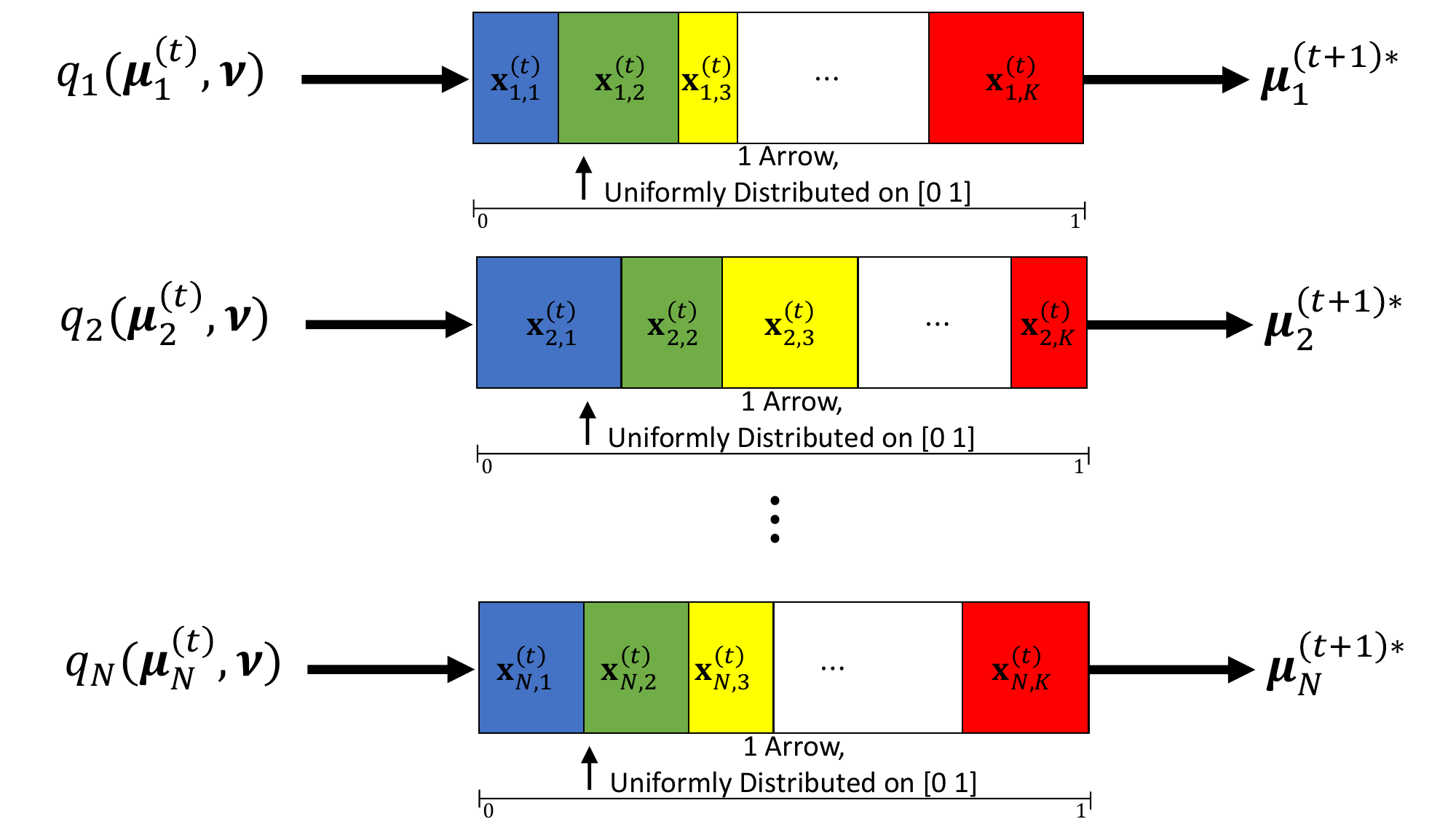}
		\caption{LR Resampling Scheme: A set of $K$ weighted samples, simulated from a proposal, is resampled to generate a new preliminary location for the next step. Each rectangle represents a sample, with its length corresponding to its normalized weight. A vertical arrow indicates a single random uniform sample drawn in each row.}
		\label{fig:lr-resampling}
	\end{figure}

	 After resampling, a new set of preliminary locations for the next iteration, $P=\{\bm{\mu}_1^{(t+1)*},\bm{\mu}_2^{(t+1)*},\cdots,\bm{\mu}_N^{(t+1)*}\}$, is obtained. There are various resampling methods in the literature \cite{li2015resampling,kuptametee2022review}. Here we use the simple multinomial LR scheme where it generates $N$ preliminary locations that are different, thereby preserving the diversity in exploring the target space (see Fig. \ref{fig:lr-resampling} for a visual illustration of how LR scheme operates).
	 
	{\it ii) Preliminary locations generation based on proposal locations}
	
\begin{figure}
		\centering
		%\begin{subfigure}[t]{0.37\linewidth}
		\includegraphics[width=0.3\textwidth]{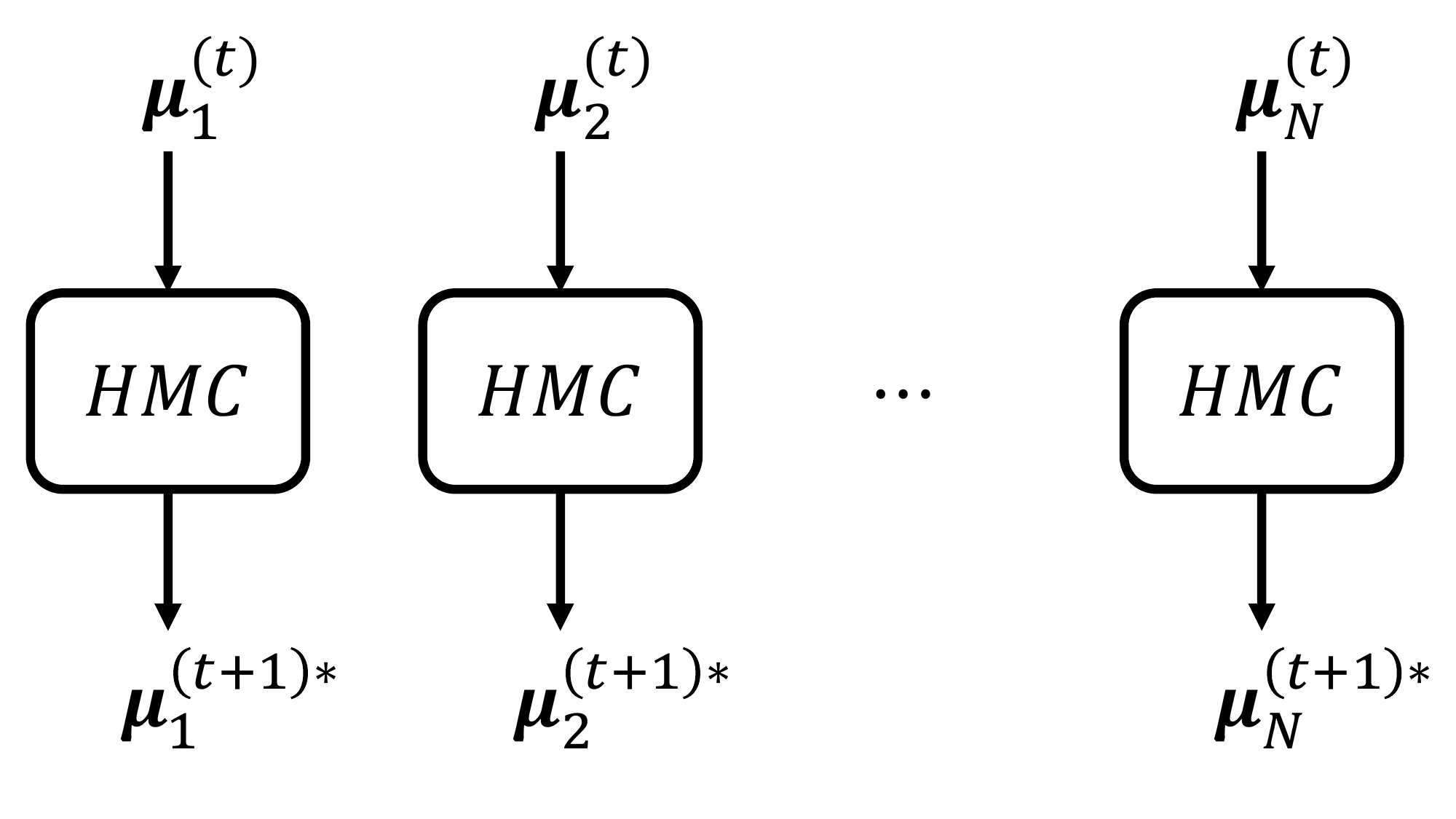}
		\caption{Preliminary locations generation based on current locations: $N$ independent HMC blocks operate in parallel to generate the new set of $N$ preliminary locations.}
		\label{fig:flowchart2}
	\end{figure}

	We generate the second set of $N$ preliminary locations from the currently available proposal locations, as shown in the top-left block of the adaptation module in Fig. \ref{fig:main_flowchart}. This process begins by running $N$ independent Hamiltonian Monte Carlo (HMC) methods in parallel. Each HMC method, represented as an HMC block in Fig. \ref{fig:flowchart2},  generates a chain with $ \bm{\mu}_n^{(1)} $ as the initial starting point. These $N$ parallel HMC blocks then produce the new set of $N$ preliminary locations, $Q=\{\bm{\mu}_1^{(t+1)*},\bm{\mu}_2^{(t+1)*},\cdots,\bm{\mu}_N^{(t+1)*}\}$, by applying one iteration of $N$ parallel chains. 
	We can consider $\widetilde{\pi}^{(t+1)*}(\bm{\mu})$ as a random measure that approximates the target distribution, i.e.,
	\begin{equation}
		%\label{eqn:randommeasure}
		\widetilde{\pi}^{(t+1)*}(\bm{\mu})=\sum_{n=1}^{N}\bar{w}_{\bm{\mu}_n^{(t+1)*}}\delta(\bm{\mu}-\bm{\mu}_n^{(t+1)*}),
	\end{equation}
	where $\bar{w}_{\bm{\mu}_n^{(t+1)*}}$ is the normalized DM weight of each $\bm{\mu}_n^{(t+1)*}$. The theoretical justification is that, following the burn-in period, the $N$ parallel HMC chains have reached convergence with the target distribution, so $\bm{\mu}_n^{(t+1)*} \sim \pi$. For more details we refer the reader to our previous HAIS method \cite{mousavi2021hamiltonian}.
	
	\textbf{Step 3(b): Final locations generation through cooperation}
	
	At this step, the goal is to determine the final proposal locations for the next iteration. Let us denote the population of all preliminary locations from Step 3(a) as
	\[
			C=\{P \cup Q\}.
	\]
	The cooperation procedure takes as input the $|C| = 2N$ preliminary locations (where $|.|$ denotes the cardinality operator) from the previous step and outputs $N$ final locations for the next iteration $(t+1)$. This process is depicted in the bottom block of the adaptation module in Fig. \ref{fig:main_flowchart}. We propose two different methods for this process: cooperation by \emph{weighted mixture model} and by \emph{resampling}.

	\paragraph{Cooperation by weighted mixture model}
	
	Here the set of $|C|$ preliminary locations is used to construct a weighted mixture model. This model serves as a proper proposal PDF to generate new samples, which are the final proposal locations for the next iteration. Let us define a weighted mixture of PDFs as
	\begin{equation}
		\label{eqn:psi_eq}
		\psi(\mu)=\sum_{i=1}^{|C|}\bar{w}_i\varphi_i(\mu|\mu_i^{(t+1)*}),
	\end{equation}
	where the $i$-th component $\varphi_i$ is a parametric kernel function with two parameters, e.g. mean and covariance in the Gaussian case. Additionally, $\mu_i$ is the $i$-th preliminary location from Step 1 serving as the location parameter of the $i$-th component. Here ${w}_i$ is the DM weight of $i$-th component $\varphi_i$, which is
	
	\begin{equation}
		\label{eqn:DM_weighting2}
		w_i=\frac{\pi(\bm{\mu}_i^{(t+1)*})}{\sum_{ j=1}^{|C|}q_j(\bm{\mu}_j^{(t+1)*};\mu_j^{(t)},\nu_j)}, \; i=1,\cdots,|C|,
	\end{equation}
	
	and then normalized by
	\[
	\bar{w}_i=\frac{{w}_i}{\sum_{j=1}^{|C|}{w}_j}, \; i=1,\cdots,|C|.
	\]
	
	$\psi(\mu)$ is a suitable proposal density for generating new samples and is updated at each AIS iteration. The theoretical justification is that, after the preliminary locations generation step, all the $\mu_i^{(t+1)*}$ have converged to the target PDF, i.e., $\mu_i^{(t+1)*} \sim \pi, \; i=1,\cdots,|C|$. As the result, $\psi(\mu)$ can be viewed as a weighted nonparametric kernel estimation of the target $\pi$.
	The iterative algorithm starts by generating one sample from $\psi(\mu)$ as a new candidate location, i.e., $\{\mu'_j\} \sim \psi(\mu)$. In the next step, $\mu'_j$  is accepted with a probability value $ \alpha_j$  then is considered as the final proposal location for the next iteration of the AIS mechanism. Alg. \ref{alg_Coop1} provides a detailed description of the proposed cooperation step.
	
	\begin{algorithm}
		%\footnotesize
		\scriptsize
		\SetAlgoLined
		\textbf{Input}: a set of $|C|$ preliminary proposal locations from the previous step, $\{\mu_i^{(t+1)*}\}_{i=1}^{|C|} \in C$ \\
		Build the weighted mixture model $ \psi(\mu)$ based on Eq. (\ref{eqn:psi_eq}) \\
		\For{$ j=1,...,N $}{
			Sample $\mu'_j \sim \psi(\mu)$ \\
			Set $ \mu_j^{(t+1)}=\mu'_j $ with probability\\
			\[
			\alpha_j=\min \left(  \frac{\pi(\mu'_j)\psi(\mu_j^{(t+1)*})}{\pi(\mu_j^{(t+1)*})\psi(\mu'_j)} \right)
			\]
			Otherwise set
			\[
			\mu_j^{(t+1)}=\mu_j^{(t+1)*}
			\]
		}
		\textbf{Output}: final set of proposal locations, $\mu_j^{(t+1)}, \; i=1,\cdots,N$ for the next iteration
		\caption{Cooperation by weighted mixture model }
		\label{alg_Coop1}
	\end{algorithm}
	
\begin{figure}[h]
		\centering
		%\begin{subfigure}[t]{0.37\linewidth}
		\includegraphics[width=0.5\textwidth]{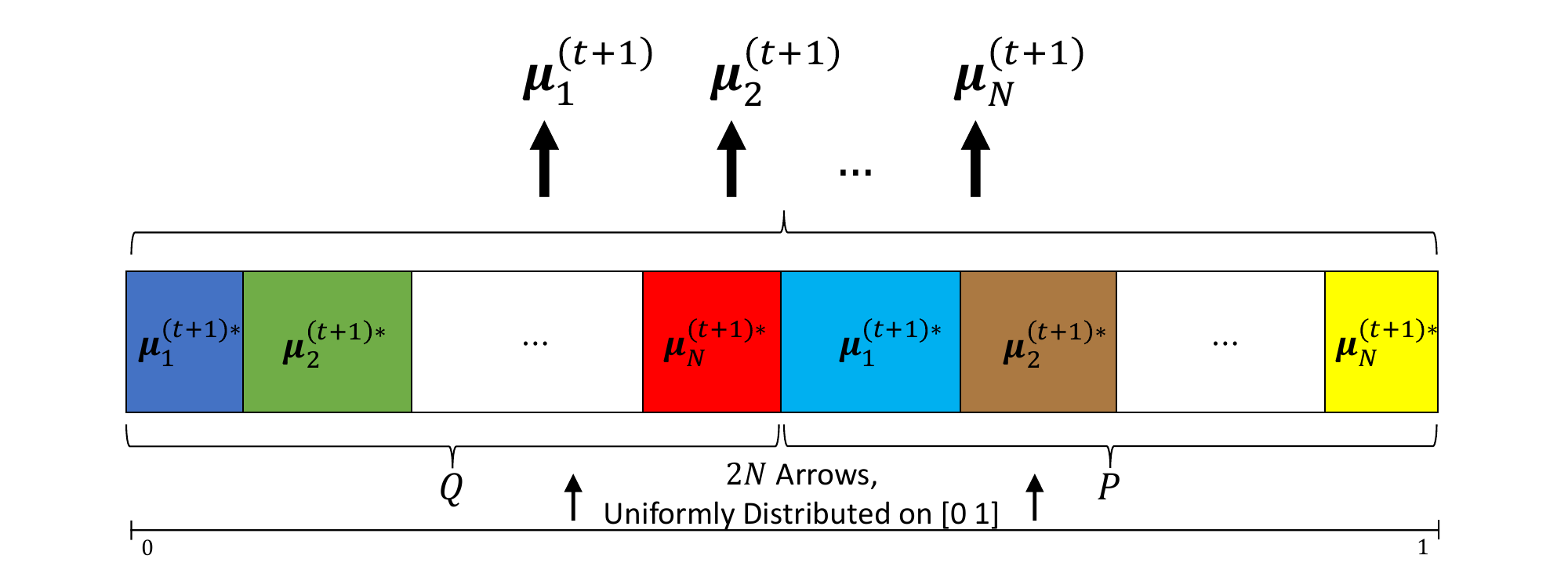}
		\caption{GR Resampling Scheme: A set of $2N$ weighted preliminary locations is resampled to generate the final location for the next iteration. Each rectangle represents a sample, with the length of the rectangle corresponding to its normalized weight. Vertical arrows indicate NN random uniform samples.}
		\label{fig:gr-resampling}
	\end{figure}

	\paragraph{Cooperation by resampling} 
	We can define $\widetilde{\pi}(\bm{\mu})$ as a random measure based on $|C|$ preliminary locations from the previous step as,
	\begin{equation}
		\widetilde{\pi}(\bm{\mu})=\sum_{i=1}^{|C|}\bar{w}_i\delta(\bm{\mu}-\bm{\mu}_i^{(t+1)*}),
	\end{equation}
	where $\bar{w}_i$ is the normalized DM weight of $i$-th preliminary location, obtained from Eq. (\ref{eqn:DM_weighting2}). The final proposal locations for the next iteration are then obtained by sampling from the random measure $ \widetilde{\pi}(\bm{\mu}) $ via global resampling (GR) \cite{elvira2017improving},
	i.e., $ {\bm{\mu}}_n^{(t+1)} \sim \widetilde{\pi}(\bm{\mu}), \; n=1,\cdots,N$. In GR, all the preliminary locations from the two sets, $P$ and $Q$, are resampled together. To better understand the idea of GR scheme see Fig. \ref{fig:gr-resampling}.
	
	After resampling, a new set of final locations for the next iteration, $ \{{\bm{\mu}}_1^{(t+1)},{\bm{\mu}}_2^{(t+1)},\cdots,{\bm{\mu}}_N^{(t+1)} \}$, is obtained, and a modified and unweighted random measure is produced as 
	\begin{equation}
		\label{eqn:randmeasure_step2-2}
		\widetilde{\pi}^{(t+1)}({\bm{\mu}})=\frac{1}{N}\sum_{n=1}^{N}\delta({\bm{\mu}}-{\bm{\mu}}_n^{(t+1)}).
	\end{equation}

\begin{figure}[h]
		\centering
		\includegraphics[width=0.5\textwidth]{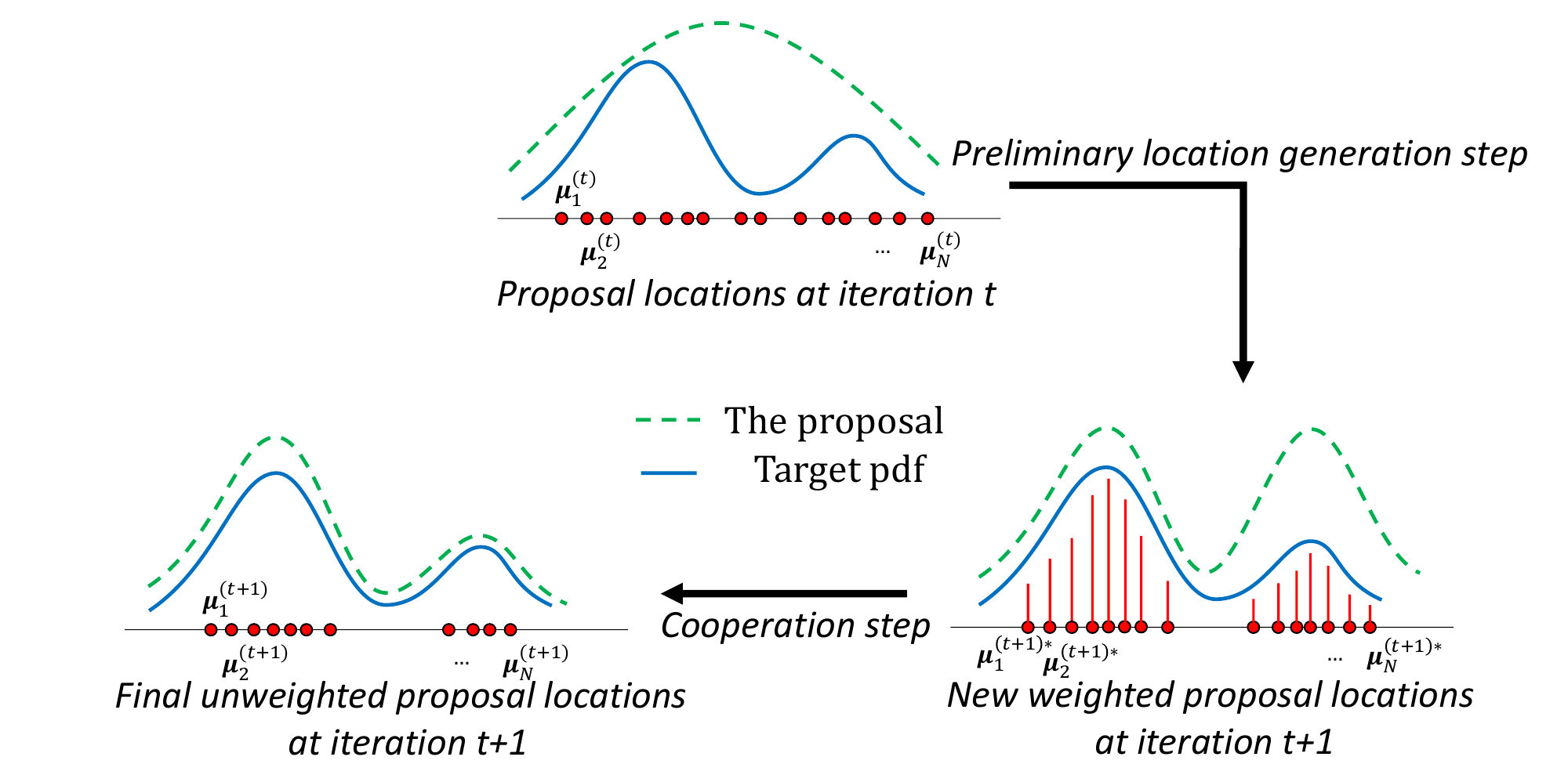}
		\caption{{The procedure of the HPMC adaptation step of the mixture of proposals (in green dashed line) in a typical
				one-dimensional bimodal target PDF (in blue solid lines). The red points are the locations of the proposals, and the red vertical lines are proportional to the importance weights.}}
		\label{fig:Fig_resampling}       
	\end{figure}

	Fig. \ref{fig:Fig_resampling} displays the proposed two-step adaptation mechanism using a univariate bi-modal target as a toy example. The new weighted proposal locations (depicted with red points as the proposal locations and red vertical lines as the corresponding weights) are obtained by performing the preliminary location generation step on the proposal locations from iteration $(t)$. However, in this case, the probability assigned to each mode in the resulting proposal PDF is not balanced, and thus it may not accurately reflect the target distribution. Subsequently, the probability of each mode is adjusted through the cooperation step, where the final unweighted proposal locations for the next iteration $(t+1)$ are generated.
	
	\subsection{Discussion}
	\label{sec_Discussion}
	In HPMC, a two-step hybrid adaptation process is presented: a preliminary location generation step and a cooperation step. Here, we discuss the key components of the HPMC algorithm:
	
	\paragraph{Preliminary location generation (local-global exploration trade-off)} 
	 In HPMC, we propose two different operations to generate preliminary locations:
	
	In the first operation, $N$ new preliminary locations are generated via down-sampling from $NK$ DM-weighted samples using a local resampling (LR) scheme. The LR approach ensures that only one preliminary location generated from each proposal location is retained. In the DM weighting applied here, the entire set of proposals in the denominator is evaluated. This interaction between proposals allows higher weights to be assigned to samples in under-populated or unexplored regions, increasing the likelihood of selecting preliminary locations from those areas. Consequently, the LR resampling with DM weights enhances the local exploration capability of the HMPC estimator, particularly in high-dimensional multi-modal target distributions. Increasing the number of samples per proposal, $K$, further improves the likelihood of discovering local relevant features of the target.
	
	In the second operation, $N$ independent HMC blocks run in parallel to generate $N$ preliminary locations for the next step. Each HMC block, with its exploratory behavior, especially in high-dimensional scenarios, acts as a \emph{free explorer} navigating the target state space. Increasing the number of these explorers enhances the global exploration capability of the estimator. Consequently, for a fixed number of target evaluations, there is a trade-off between promoting local exploration (by increasing $K$) and enhancing global exploration (by increasing $N$). By leveraging both weighted samples and location parameters, HPMC reduces the number of iterations required to achieve the target estimate, leading to faster convergence.
	
	\paragraph{Cooperation} 
	In this step, information is exchanged among the preliminary locations to redirect the lost explorers toward the optimal path based on the target PDF. Cooperation is performed using two different methods, both applying the DM weighting approach to assign weights to the preliminary locations from the previous step. The DM weighting enforces a repulsive behavior among the locations by using the complete mixture of all proposals in the denominator, as discussed earlier. In other words, locations that are less covered by the remaining $|C|-1$ locations receive higher weights, enhancing the exploratory behavior of the cooperation step. In the resampling method, the GR scheme with global DM weights is used to terminate lost explorers in irrelevant regions of the target, while replicating more promising ones—those with higher global weights.
	\section{Numerical Examples}
	\label{sec_Experiments}
	In this section, we evaluate the efficiency of the proposed method through three experiments and compare its performance with existing methods in the literature. The first experiment considers a two-dimensional multi-modal target as a toy example. The second experiment evaluates a high-dimensional bimodal target probability density function (PDF). Finally, the third experiment involves a challenging benchmark problem for Monte Carlo (MC) algorithms: sampling from a banana-shaped distribution at different dimensions. We also compare the computational complexity of the proposed method with other exiting algorithms in terms of $K$ (number of samples generated from each proposal PDF), $N$ (number of proposals PDFs), and $T$ (number of iterations). 
	
	In all experiments, the number of iterations, $ T $, for each method is set such that they have the same total number of target evaluations, $ E=2\times 10^5 $. We compute the mean squared error (MSE) in the estimation of $ \mathbb{E}[\x] $ and the normalizing constant $Z$ of the target. The results are averaged over 200 Monte Carlo simulations.
	
	We compare the proposed hybrid population Monte Carlo (HPMC) techniques with 
	\begin{itemize}
		\item  standard PMC \cite{cappe2004population};
		\item DM-PMC, GR-PMC and LR-PMC methods \cite{elvira2017improving}; 
		\item AMIS \cite{cornuet2012adaptive} with one single proposal PDF ($N=1$); %so we choose $K=500$;
		\item PI-MAIS \cite{martino2017layered} where for the upper layer, we also consider Gaussian PDFs $ \varphi_n(\x|\bm{\mu}_n,\Lambda_n) $ where covariance matrices $\Lambda_n=\lambda^2I_2$ with $\lambda \in \{5,10\}$;
		\item HAIS \cite{mousavi2021hamiltonian} with $\epsilon \in \{5,10\}$ and $L=50$.
	\end{itemize}
	
	\subsection{Toy example}
\begin{figure*}[h!]
		\centering
		\subfigure[GR-PMC (iteration 1)]{
			\includegraphics[scale=0.25]{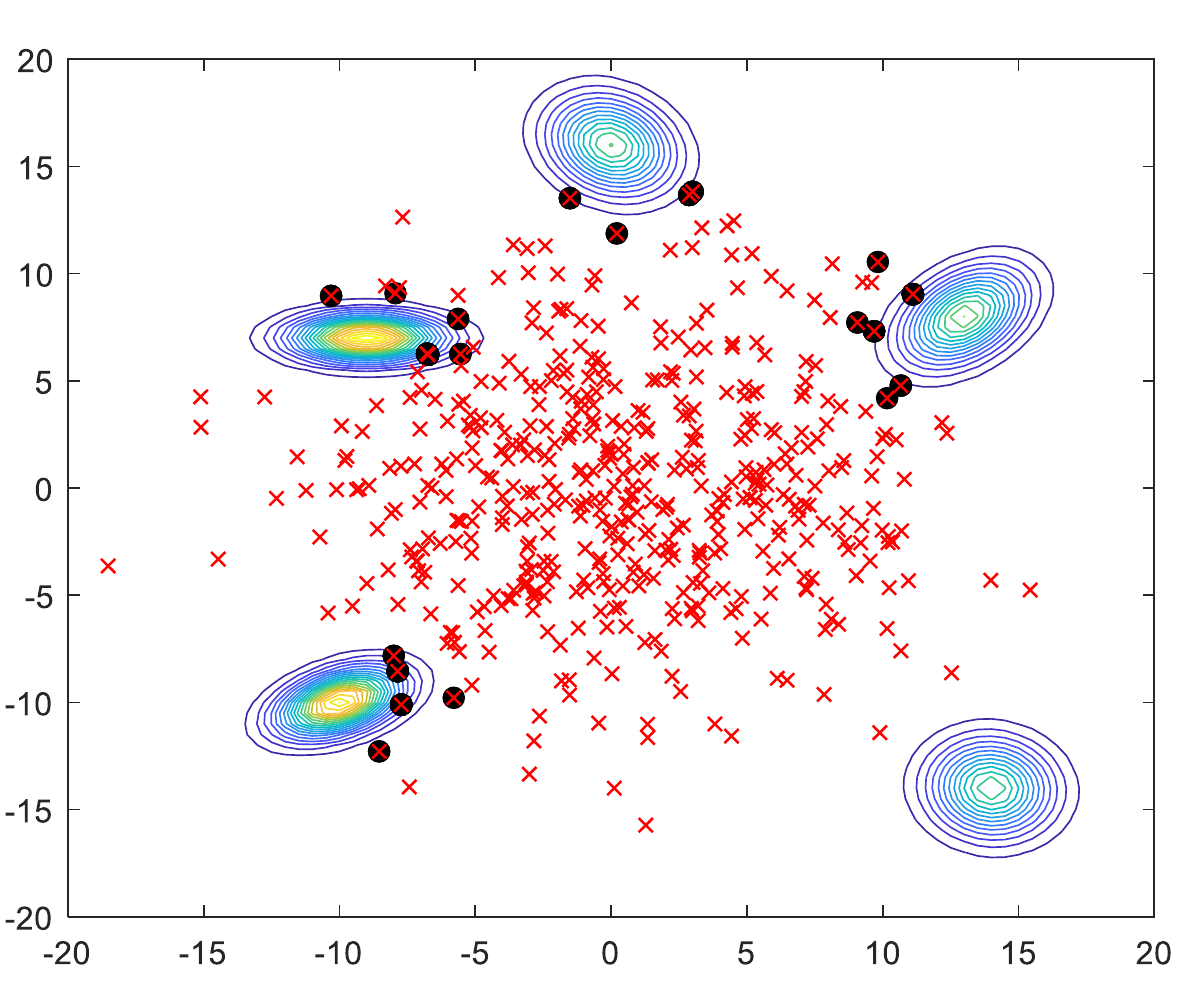}
		} \quad
		\subfigure[GR-PMC (iteration 2) ]{
			\includegraphics[scale=0.25]{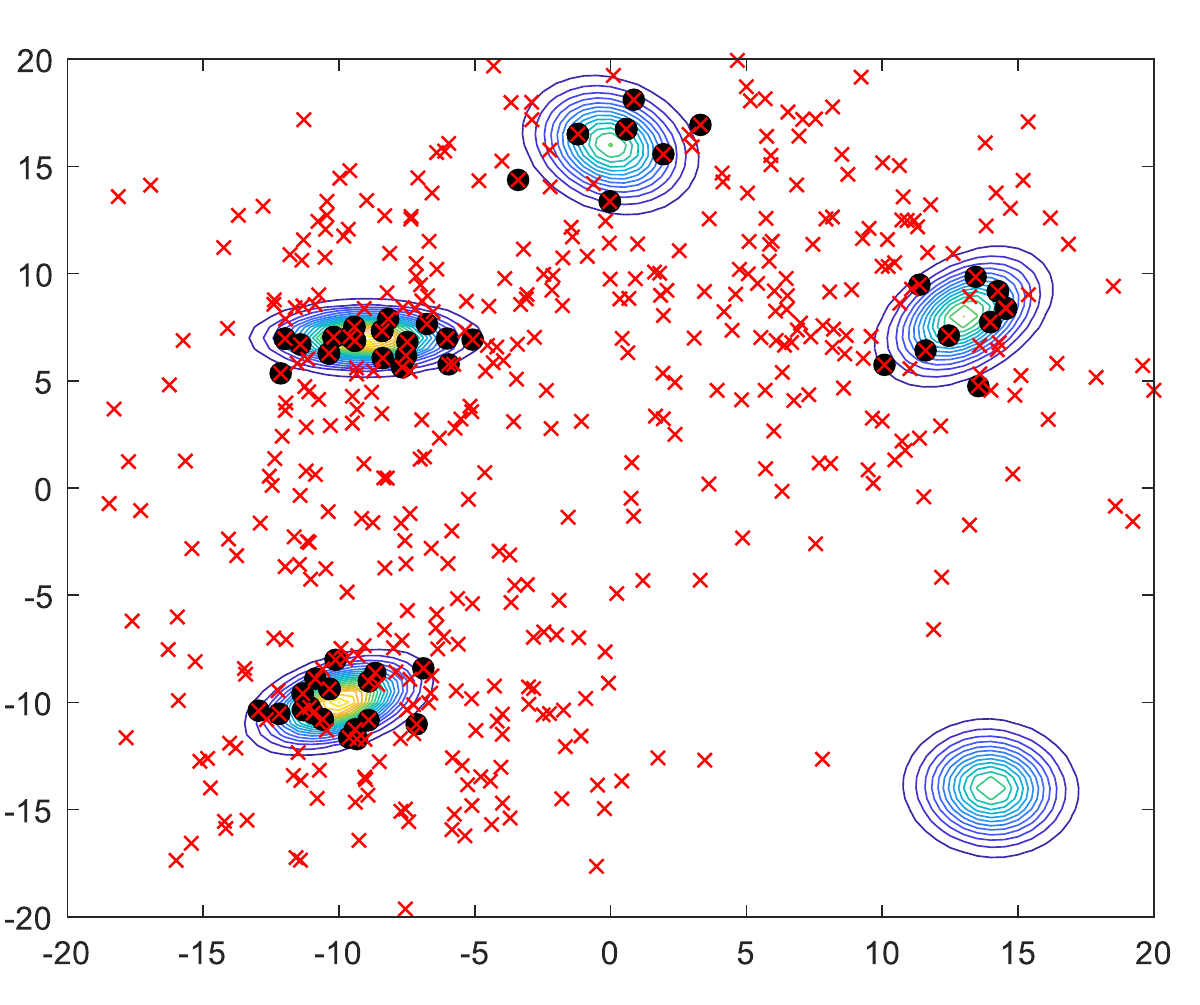}
		} \quad
		%\centering
		\subfigure[GR-PMC (iteration 3)]{
			\includegraphics[scale=0.25]{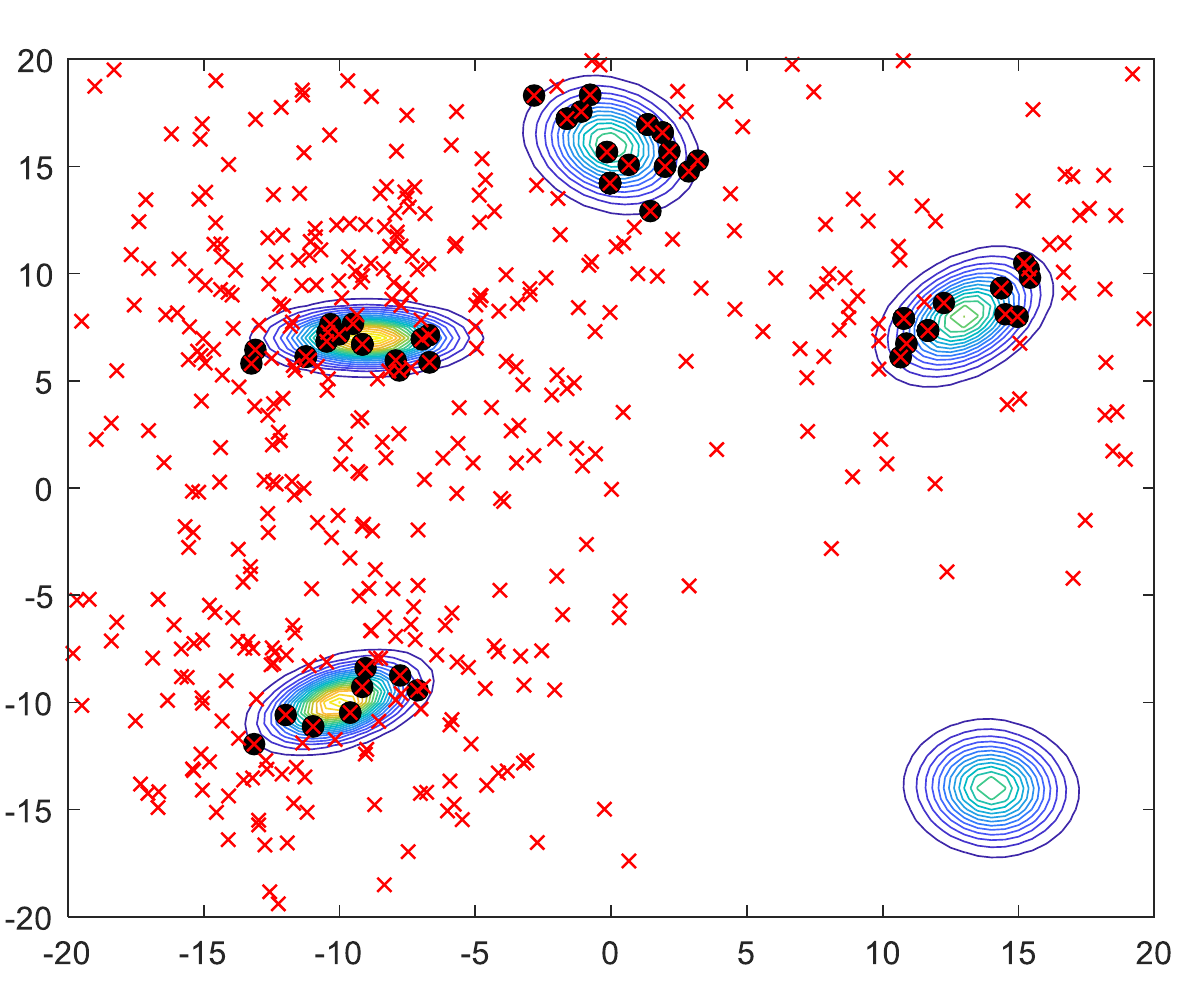}
		}
		\centering
		\subfigure[LR-PMC (iteration 1)]{
			\includegraphics[scale=0.25]{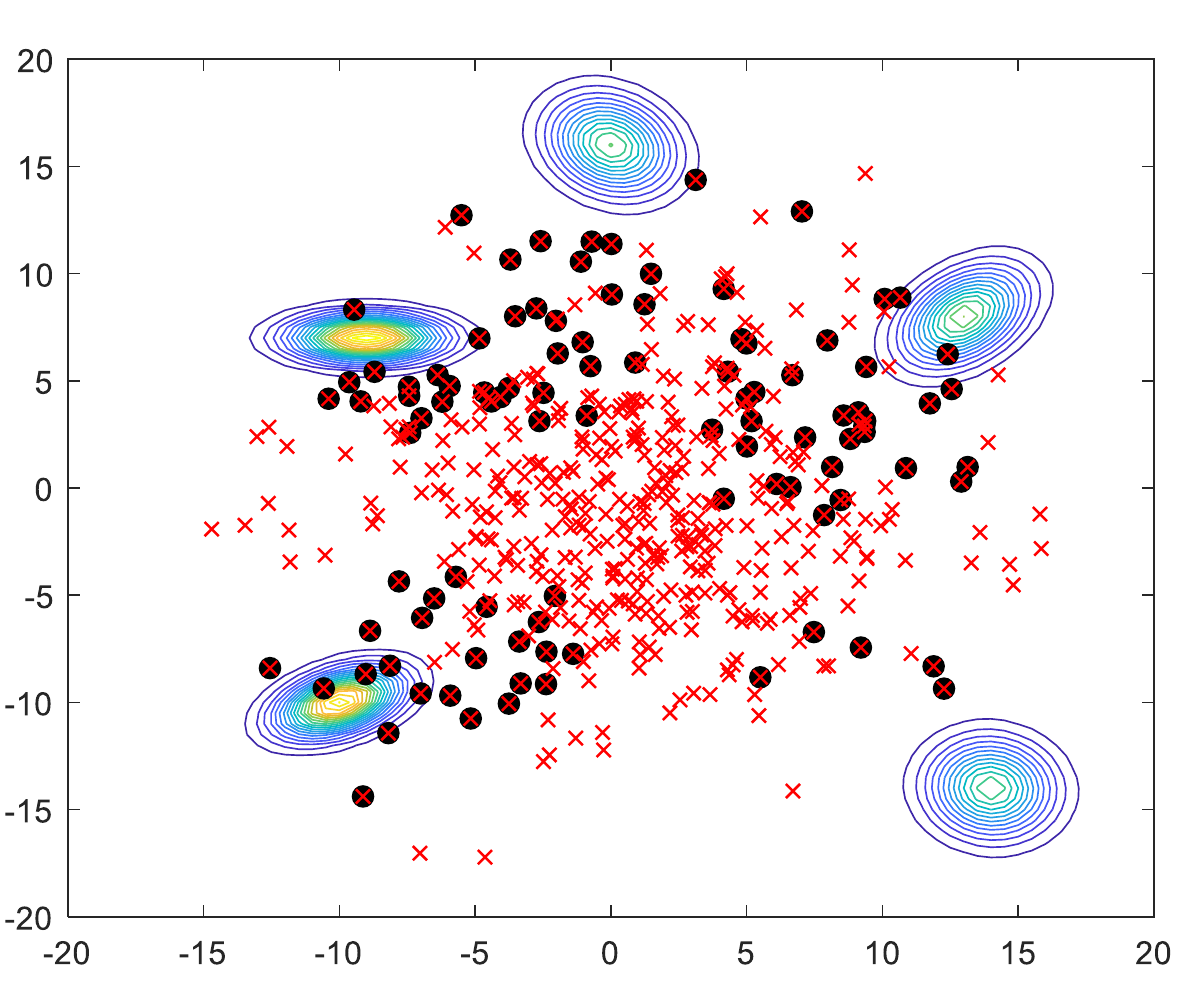}
		} \quad
		\subfigure[LR-PMC (iteration 2)]{
			\includegraphics[scale=0.25]{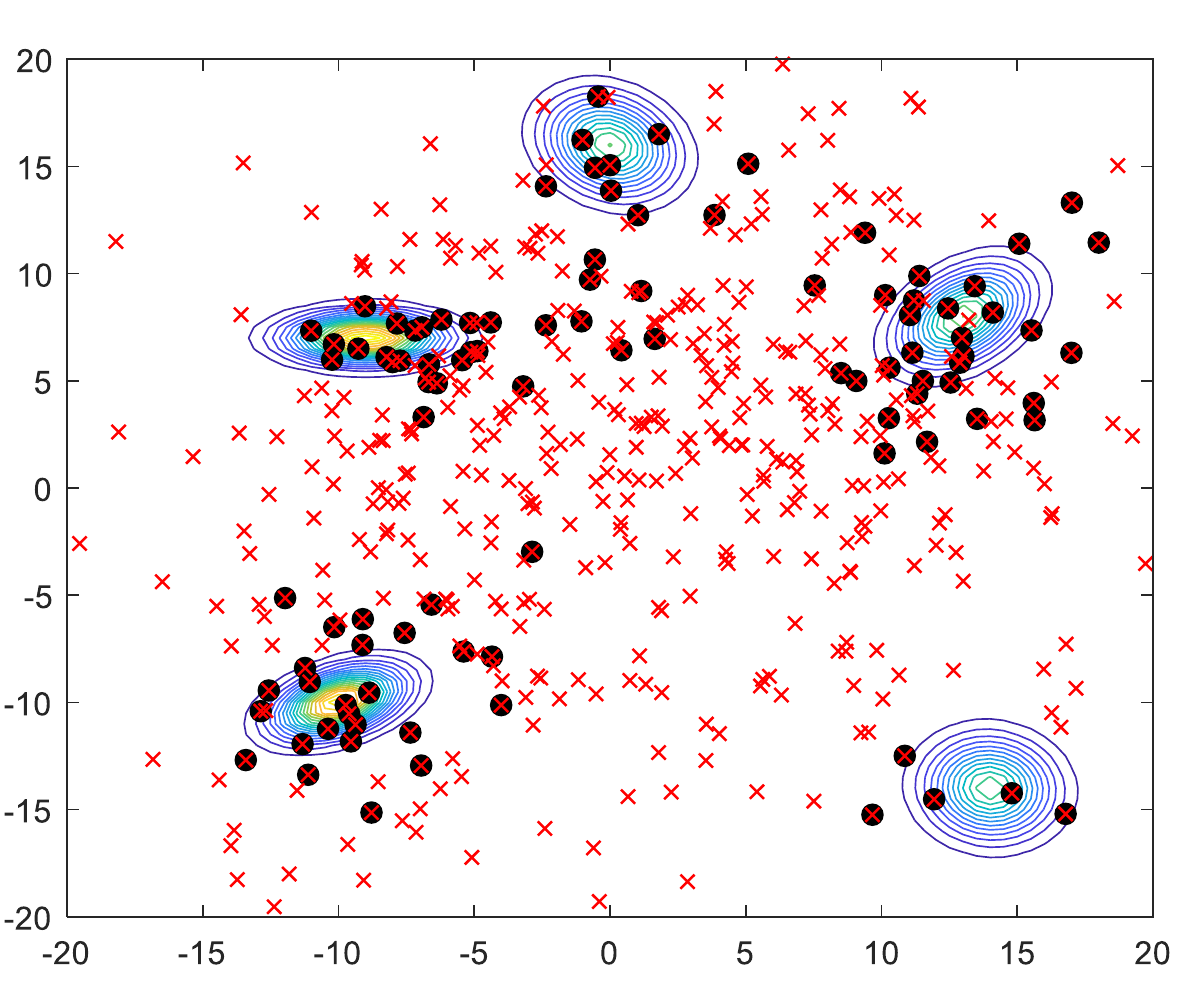}
		} \quad
		\subfigure[LR-PMC (iteration 3)]{
			\includegraphics[scale=0.25]{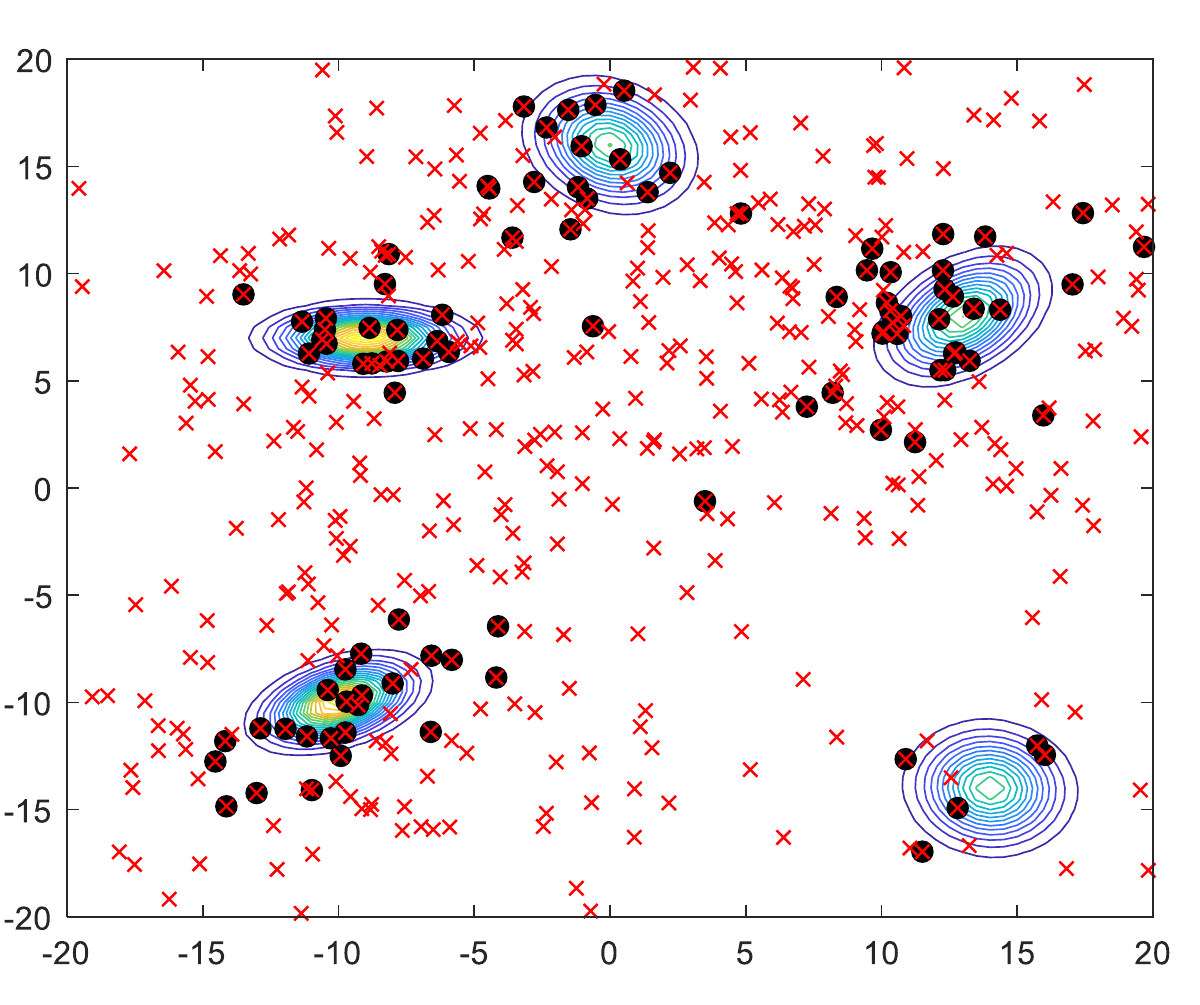}
		} \quad
		\centering
		\subfigure[PI-MAIS (iteration 1)]{
			\includegraphics[scale=0.25]{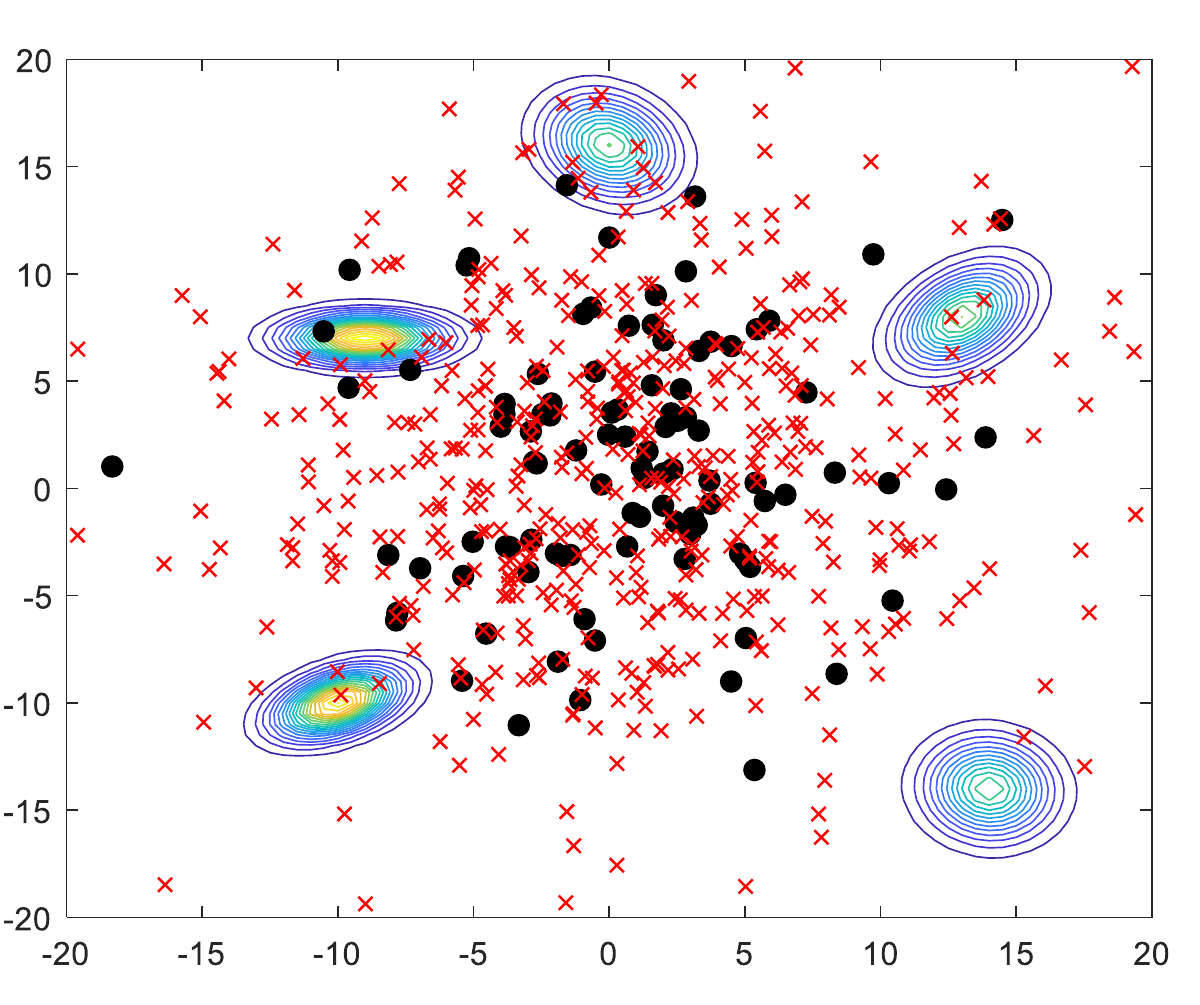}
		} \quad
		\subfigure[PI-MAIS (iteration 2)]{
			\includegraphics[scale=0.25]{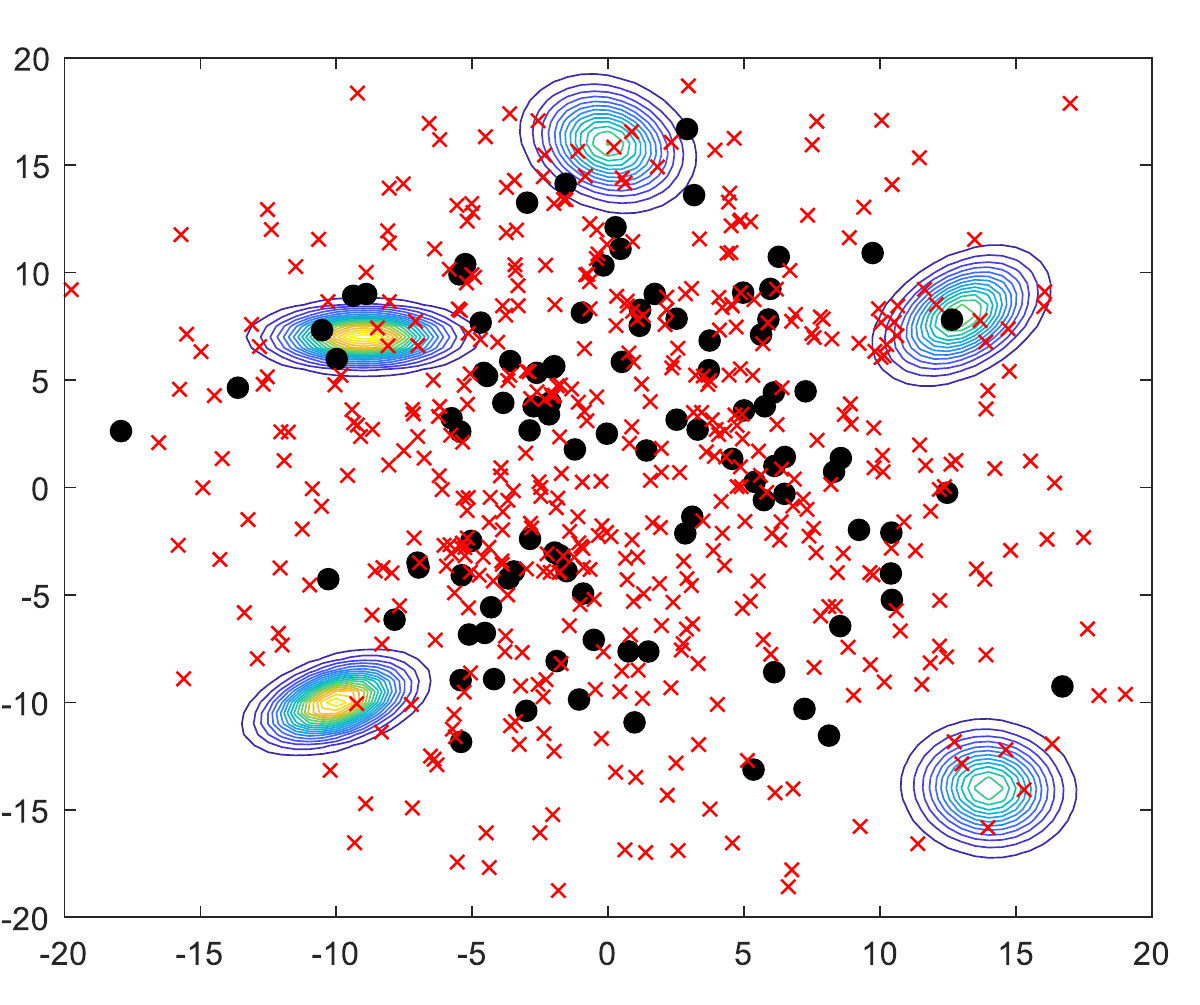}
		} \quad
		\subfigure[PI-MAIS (iteration 3)]{
			\includegraphics[scale=0.25]{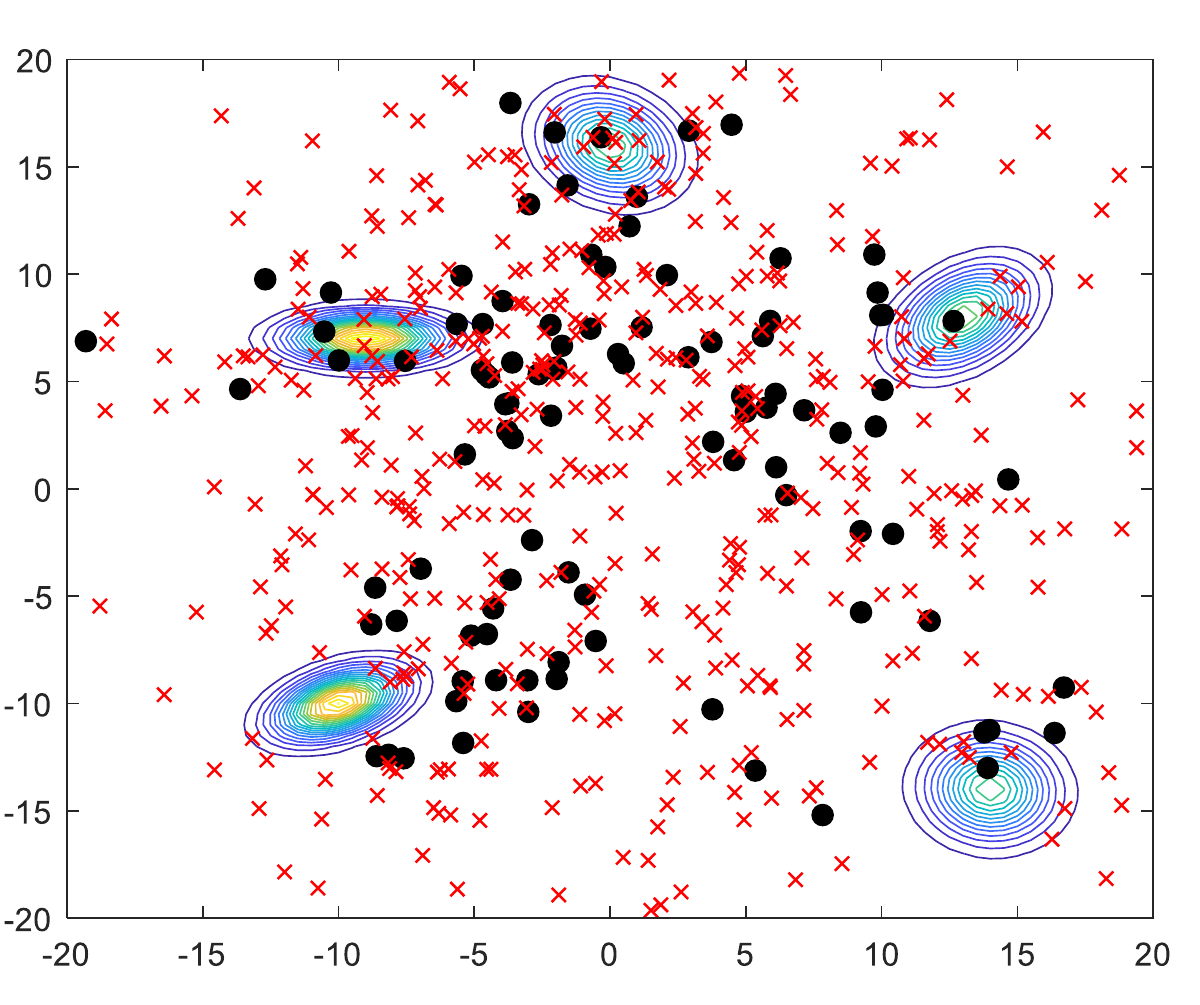}
		} \quad
		\centering
		\subfigure[HPMC (iteration 1)]{
			\includegraphics[scale=0.25]{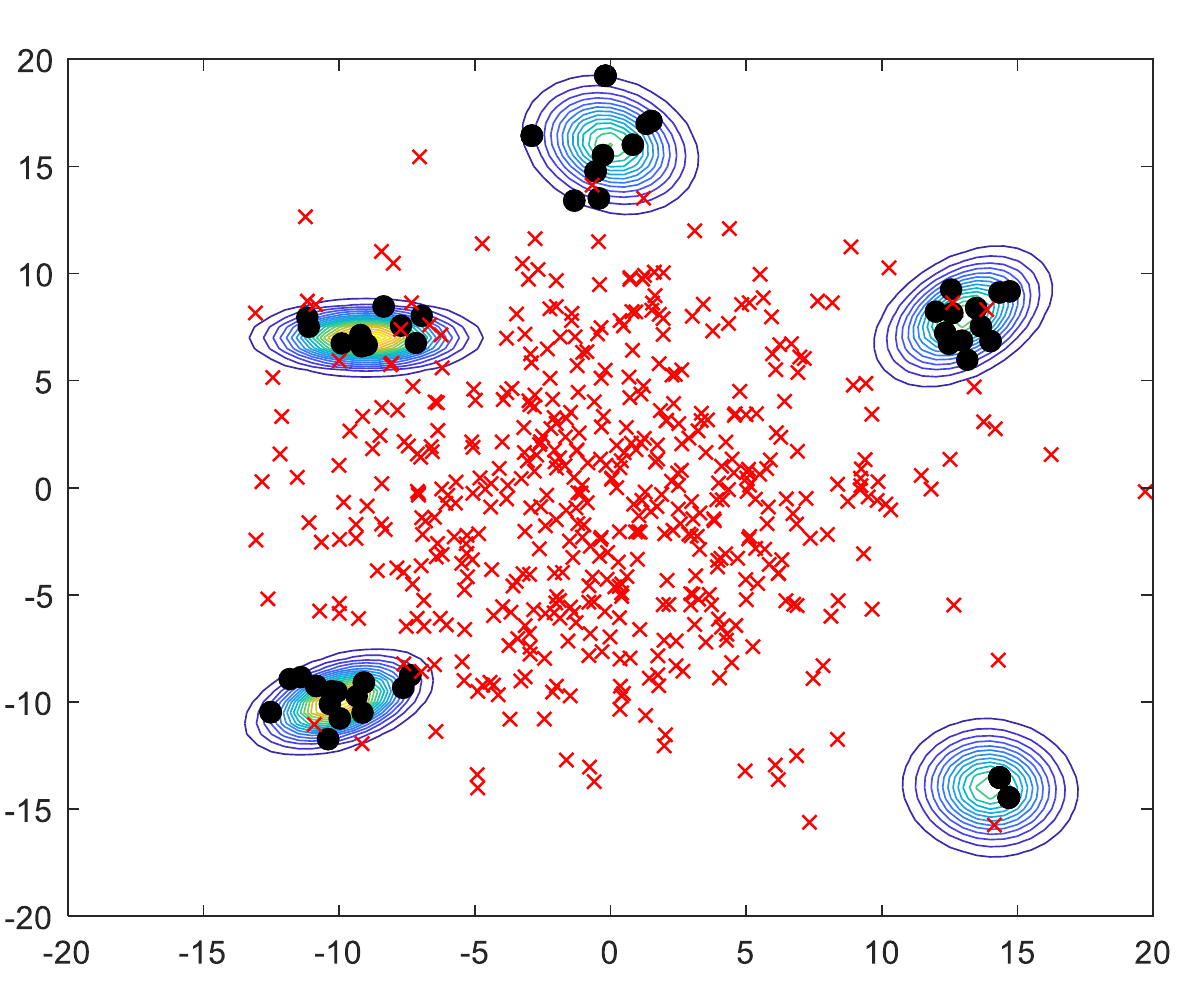}
		} \quad
		\subfigure[HPMC (iteration 2)]{
			\includegraphics[scale=0.25]{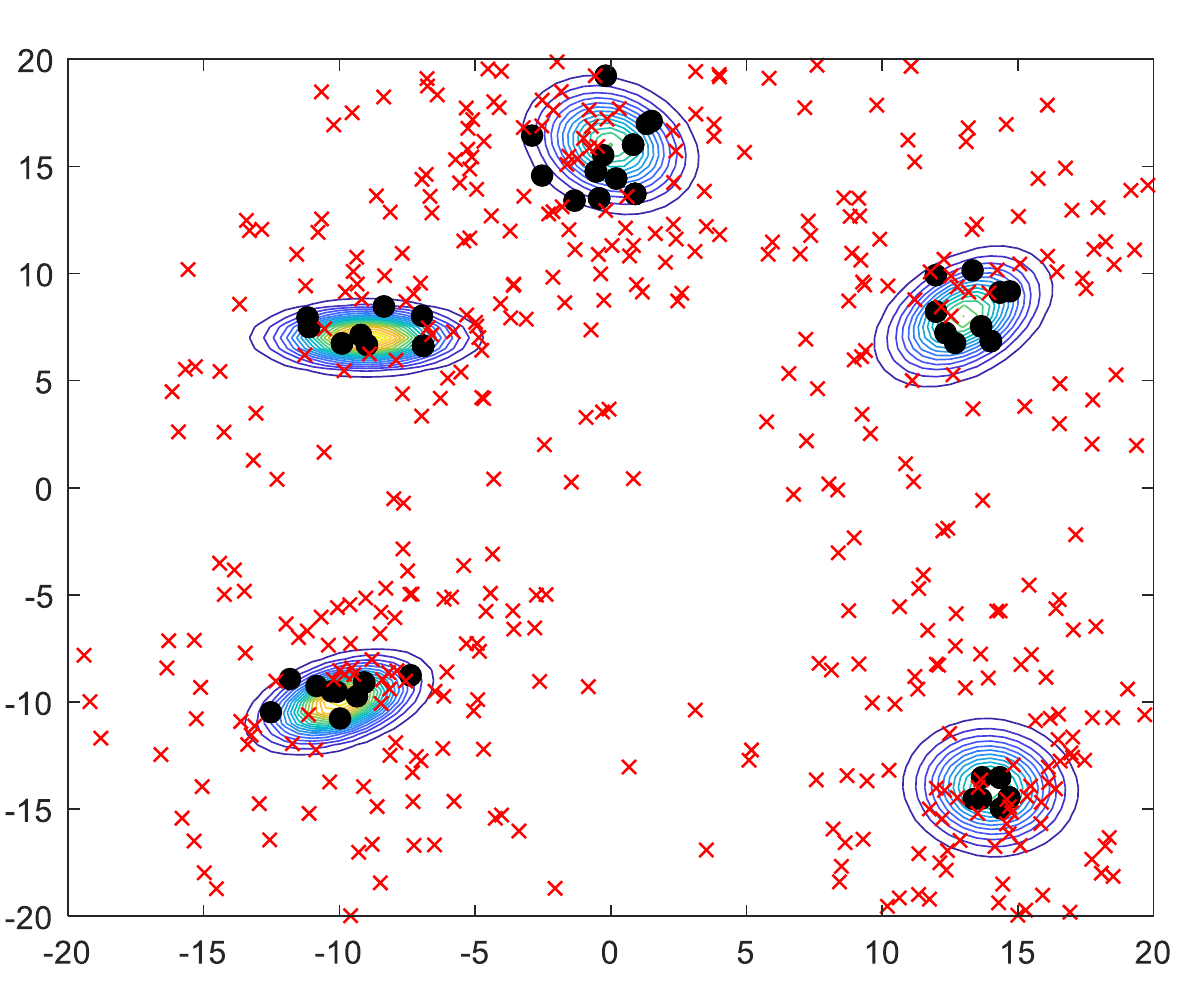}
		} \quad
		\subfigure[HPMC (iteration 3)]{
			\includegraphics[scale=0.25]{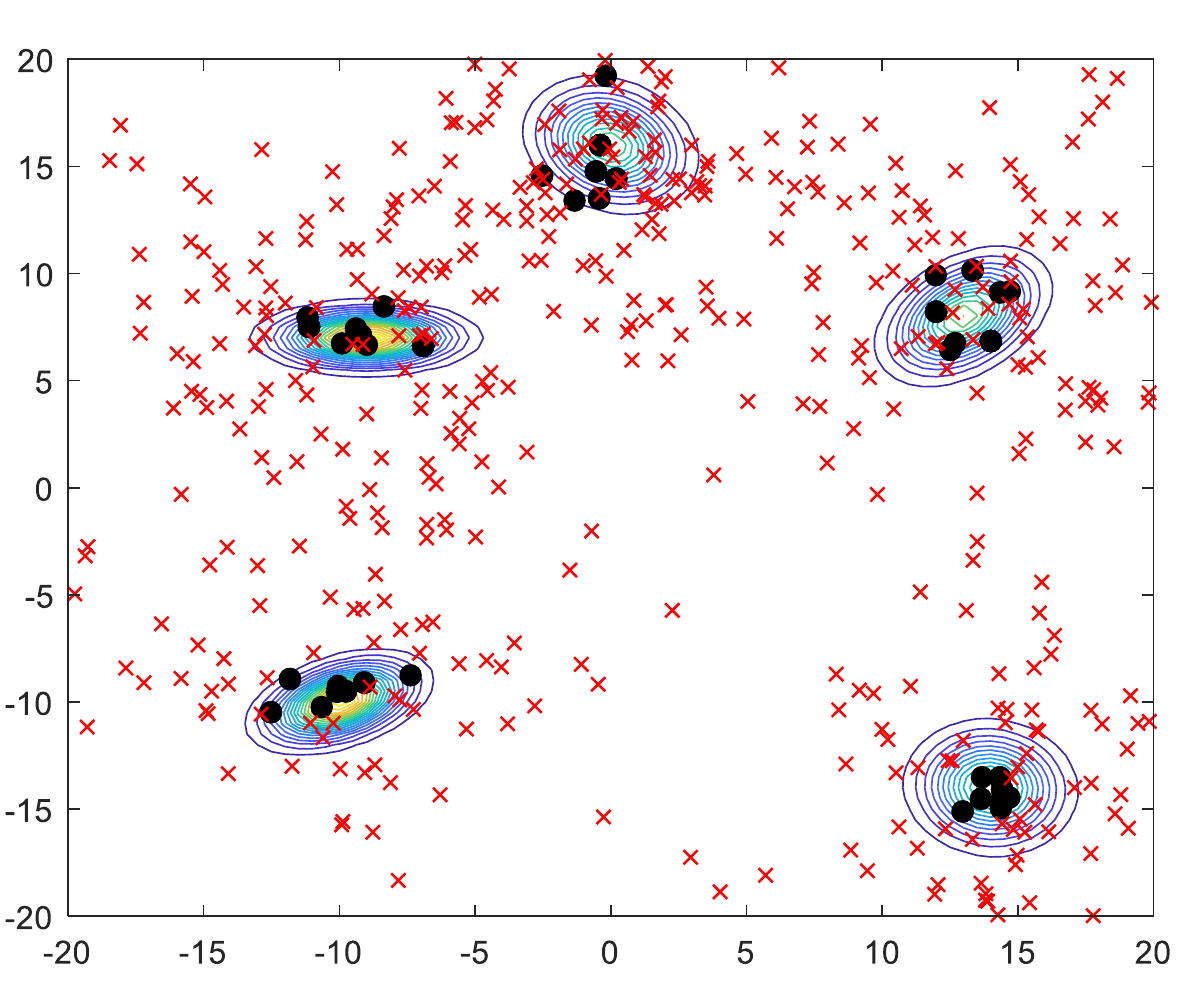}
		} \quad
		\caption{Toy example: Evolution of proposal locations (black circles) and samples (red crosses) along three iterations for GR-PMC, LR-PMC, PI-MAIS, and HPMC using $N=100$ and $K=5$. The multi-modal target density with five modes is shown in contours in 2D plots. }
		\label{figure_toy_example}
	\end{figure*}

	We consider a two-dimensional example to examine how different methods behave along iterations. The target is a mixture of five bivariate Gaussian PDFs 
	\begin{equation}
		\pi(\x)=\frac{1}{5}\sum_{i=1}^{5} \mathcal{N}(\bm{\mu}_i,\,\bm{\Sigma}_i), \; \x\in \mathbb{R}^{2},
	\end{equation} 
	where the vector of means are $ \bm{\mu}_1=[-10, -10]^T$, $ \bm{\mu}_2=[0, 16]^T$, $ \bm{\mu}_3=[13, 8]^T$, $ \bm{\mu}_4=[-9, 7]^T$, $ \bm{\mu}_5=[14, -14]^T$, and the covariance matrices are $\bm{\Sigma}_1=[2, 0.6;0.6, 1]^T$, $\bm{\Sigma}_2=[2, -0.4;-0.4, 2]^T$, $\bm{\Sigma}_3=[2, 0.8;0.8, 2]^T$, $\bm{\Sigma}_4=[3, 0;0, 0.5]^T$, and $\bm{\Sigma}_5=[2, -0.1;-0.1, 2]^T$. This is a challenging example due to difficulty in discovering all five different modes, particularly if one selects a bad initialization for proposals. We use isotropic Gaussian proposals with locations within the square $[-4,4]\times[-4,4]$, i.e., $\{\bm{\mu}_n^{(1)}\}_{n=1}^N  \sim U\left([-4,4]\times[-4,4]\right)$, where all of the modes in the target PDF are out of the initialization area. We also use the same isotropic covariance matrices for all the proposals, $\{\bm{\Sigma}_n\}_{n=1}^N=\sigma^2I_2$ with $\sigma=5$. We set $N=100$ proposals, and $K=5$ sample per proposal. We run GR-PMC and LR-PMC methods \cite{elvira2017improving}, PI-AMIS \cite{martino2017layered}, and HPMC with the same setting for a fair comparison. Fig. \ref{figure_toy_example} shows the graphical evolution through the first three iterations of the proposal locations (black circles) and generated samples (red crosses). In HPMC, the proposal locations rapidly discover all five modes in the first iteration, and an approximately uniform distribution of the proposal locations around all the target modes is achieved in the next iterations. So an accurate target estimation is obtained after a few iterations. PI-AMIS almost finds all five modes after three iterations, showing slow convergence, as the proposals are located between the initial positions and the centers of the target modes. One limitation of PI-AMIS is the well-known random-walk behavior of the MH that makes the convergence of the Markov chain inefficient. Both LR-PMC and GR-PMC are failed to estimate the target accurately. The local resampling technique in LR-PMC maintains the diversity of the proposal population, leading to the discovery of all target modes. However, it lacks global approximation because the proposal locations are unevenly distributed. In GR-PMC, after three iterations, the proposals are almost distributed around four modes while one mode is completely missed.
	
	\subsection{Bimodal target distribution in high dimension}
	In this section, we consider sampling from a high-dimensional bi-modal target PDF as a challenging problem. The target is a mixture of two Gaussians, i.e., $\pi(\x)=\sum_{i=1}^{2} \frac{1}{2}\mathcal{N}(\bm{\mu}_i,\,\bm{\Sigma}_i)$. Here $ \x\in \mathbb{R}^{20} $, $ \bm{\mu}_i=[{\mu}_{i,1},{\mu}_{i,2},...,{\mu}_{i,20}]^T $, and $ \bm{\Sigma}_i=c\bI_{20} $, for $ i\in \{1,2\} $ where $ \bI_{20} $ is an identity matrix of dimension  $ 20$. We choose $ {\mu}_{1,d}=8 $ and $ {\mu}_{2,d}=-8 $ for all dimensions and also we set $ c=5 $. Multi-modal settings are challenging, and in this example the two modes are distant, which over-complicates the exploration process at high dimension. Moreover, the initial starting points play a crucial role in exploring the target. Here, we consider a cold-start scenario to evaluate robustness and adaptation capabilities. For simplicity, we assume the proposal densities are Gaussian PDFs with uniformly selected initial means, i.e., $ \bm{\mu}_n \sim U([-4\times 4]^{d_x}) $ for $ n=1,...,N $, where none of the modes of the target fall within this area. We use the same isotropic covariance for all of proposals, $ \bm{\Sigma}= \sigma^2\bI_{20} $ with $ \sigma\in\{1,2,5,10,20\} $. 
	
	We compute MSE in the estimation of the mean and the normalizing constant of the target, which can be find analytically as $ \mathbb{E}[\x]=0 $ and $ Z=1$. The MSE in the estimation of $ \mathbb{E}[\x]$ is averaged over all 20 components. All algorithms are simulated with a fixed number of total generated samples in $T$ iterations, i.e., $NK=500$. We test for two scenarios, once we choose $N=100$ and $K=5$ and in the second scenario we select $N=250$ and $K=2$. The simulation results from each method are summarized in Table \ref{table1_MSE_E} in terms of MSE in the estimation of $\mathbb{E}[\x]$ and $Z$ for different configurations of $N$, $K$ and various range of $\sigma$ values.
	
	First, note that most algorithms result in very large MSE values. These situations often occur when one or both modes fail to be discovered. When both modes are missed, the estimation of normalizing constant of the target is $ \widehat{Z}\approx 0 $ which corresponds to a MSE$\approx 1$, as it happens in several settings. 
	Second, we observe that the two proposed schemes outperform the other methods. According to this table, HPMC algorithm with cooperation by resampling generally achieves the smallest MSE values for the mean and normalizing constant estimation, (MSE$_{\mathbb{E}[\x]}=7.7889$ and MSE$_Z=0.0009$), where $N=250$, $K=2$, and $\epsilon=5$. It is evident that increasing $N$ has a greater impact on MSE reduction than increasing $K$. The intuition behind this is that increasing the number of proposals, with each proposal acting as an independent explorer of the target space, enhances the global exploration capability of the sampling method. We note that the optimal value of $ \sigma $, which yields the smallest MSE, depends on the scale parameter of the target (here it is $ c=5 $).
	
	\begin{table*}[t!]
		\caption{MSE in the approximation of the mean and normalizing constant of multi-modal target distribution. The best results for each value of $ \sigma $ are boldfaced}
		\label{table1_MSE_E}
		\centering
		\begin{tabular}{c c c c c c c c c c c} \hline \hline
			{Method} & \multicolumn{2}{c}{$ \sigma=1 $} & \multicolumn{2}{c}{$ \sigma=2 $} & \multicolumn{2}{c}{$ \sigma=5 $}& \multicolumn{2}{c}{$ \sigma=10 $} & \multicolumn{2}{c}{$ \sigma=20 $}     \\  \cline{2-11}
			&   $ \mathbb{E}[\x] $ &   $ Z $    &  $ \mathbb{E}[\x] $  & $ Z $      & $ \mathbb{E}[\x] $   & $ Z $ & $ \mathbb{E}[\x] $   & $ Z $ & $ \mathbb{E}[\x] $   & $ Z $        \\ \hline \hline
			Standard PMC \\
			$N=100, K=1$ & 70.1248 & 1 & 71.0234 & 1 & 69.7048 & 1 & 78.3254 & 1 & 152.3154 & 1\\ \hline 
			DM-PMC \\
			$N=100, K=1$ & 66.2487 & 1 & 67.0015 & 0.9999 & 66.3021 & 1 & 69.3215 & 1 & 149.0478 & 1\\ \hline 
			LR-PMC  \\
			$N=100, K=5$ & 21.5680 & 0.9974 & 26.7301 & 0.0998 & 65.2457 & 0.9997 & 65.0219 & 1 & 137.2168 & 1 \\
			$N=250, K=2$ & 20.1789 & 0.9545 & 24.8894& 0.0754 & 46.2145 & 0.9632 & 53.0985 & 1 & 131.0245 & 1\\ \hline 
			GR-PMC  \\
			$N=100, K=5$ & 64.9203 & 0.9418 & 64.1398 & 0.3126 & 65.7060 & 0.3412 & 83.9430 & 1 & 143.4574 & 1\\
			$N=250, K=2$ & 61.0387 & 0.9105 & 62.0985 & 0.2985 & 56.2015 & 0.3101 & 71.0215 & 1 & 140.2015 & 1 \\ \hline
			AMIS  \\
			$N=1, K=500$ & 80.2054 & 1 &  79.0548 & 0.9998 & 82.0195 & 0.9996 & 102.0215 & 1 & 159.0598 & 1\\ \hline
			PI-MAIS\\
			$N=100, K=5$ \\
			$\lambda=5$& 46.1009 & 1 & 38.9915 & 0.4531 & 36.6260 & 0.5421 & 65.5964 &  0.9389 & 135.5524 & 1 \\
			$\lambda=10$ & 60.5114 & 1 & 60.3086 & 1 & 53.2838 & 0.9878 & 61.7771 & 0.9946 & 130.5375 & 1 \\
			$N=250, K=2$\\
			$\lambda=5$ & 41.1068 & 0.9568 & 37.0658 & 0.3057 & 31.0287 & 0.3924 & 61.0548 & 0.9702 & 131.2050 & 1 \\
			$\lambda=10$ & 43.5145 & 0.9328 & 45.0916 & 0.8576 & 35.0367 & 0.5205 & 62.0348 & 0.9905 & 132.0957 & 1 \\ \hline
			HAIS \\ 
			$N=100, K=5$ \\
			$ \epsilon  = 5 $& 41.0925 & 0.8649 &  17.5741 & 0.0201 & 13.1954 & 0.0031 & 58.0607 & 0.8573 & 109.0967 & 0.9899\\
			$ \epsilon  = 10 $& 42.7641 & 0.8828 & 17.3210 & 0.0162 & 12.8675 & 0.0016 &  61.5772 & 0.9823 & 117.7326 & 1 \\
			$N=250, K=2$ \\
			$ \epsilon  = 5 $& 32.0568 & 0.7565 & 15.0248 & 0.0169 & 9.0249 & 0.0013 & 41.0254 & \textbf{0.7535} & \textbf{99.0320} & 0.9855\\
			$ \epsilon  = 10$& 34.9875 & 0.7575 & 17.0254 & 0.0113 & 11.0587 & 0.0083 & 48.0548 & 0.7814 & 104.0032 & 0.9917\\ \hline
			HPMC \\
			(cooperation by mixture model) \\
			$N=100, K=5$ \\
			$ \epsilon  = 5 $& 39.0215 &0.8816 &17.0248 & 0.0209&16.2054 & 0.0990&55.2213 &0.8641 &113.0558&0.9888\\
			$ \epsilon  = 10 $& 40.0215 & 0.8791& 19.8332 & 0.0193&15.9683 & 0.0117&60.3506 & 0.9401&117.9220 &0.9990\\
			$N=250, K=2$ \\
			$ \epsilon  = 5 $& 33.3467 &0.7639 &15.2418 & 0.0161&10.5921 & 0.0088&49.6691 & 0.8857&110.9152 &0.9989\\
			$ \epsilon  = 10 $&34.0122& 0.7501&16.9735& 0.0126& 11.9617& 0.0107&50.1877& 0.7819&112.2142 & 0.9901 \\ \hline
			HPMC \\
			(cooperation by resampling) \\
			$N=100, K=5$ \\
			$ \epsilon  = 5 $& 21.7384&0.7508&15.2681&0.0217&8.4948&0.0018&45.7573&0.8820&110.3492&0.9809 \\
			$ \epsilon  = 10 $&22.7792 &0.7444&17.5407&0.0170&9.4434&0.0015&44.3853&0.8925&112.14.15&0.9878 \\
			$N=250, K=2$ \\
			$ \epsilon  = 5 $& \textbf{19.3494}&\textbf{0.6211}&\textbf{14.5010}&0.0127&\textbf{7.7889}&\textbf{0.0009}& \textbf{40.9551}&0.7591&100.1874&\textbf{0.9790}\\
			$ \epsilon  = 10 $&21.7117&0.6496&15.6816&\textbf{0.0111}&9.8156&0.0011&43.1310&0.7747&101.8411&0.9784\\ \hline
		\end{tabular}
	\end{table*}
		
	\subsection{High-dimensional banana-shaped target distribution}
\begin{figure}
		\centering
		\begin{subfigure}
			\centering
			\includegraphics[width=0.79\linewidth]{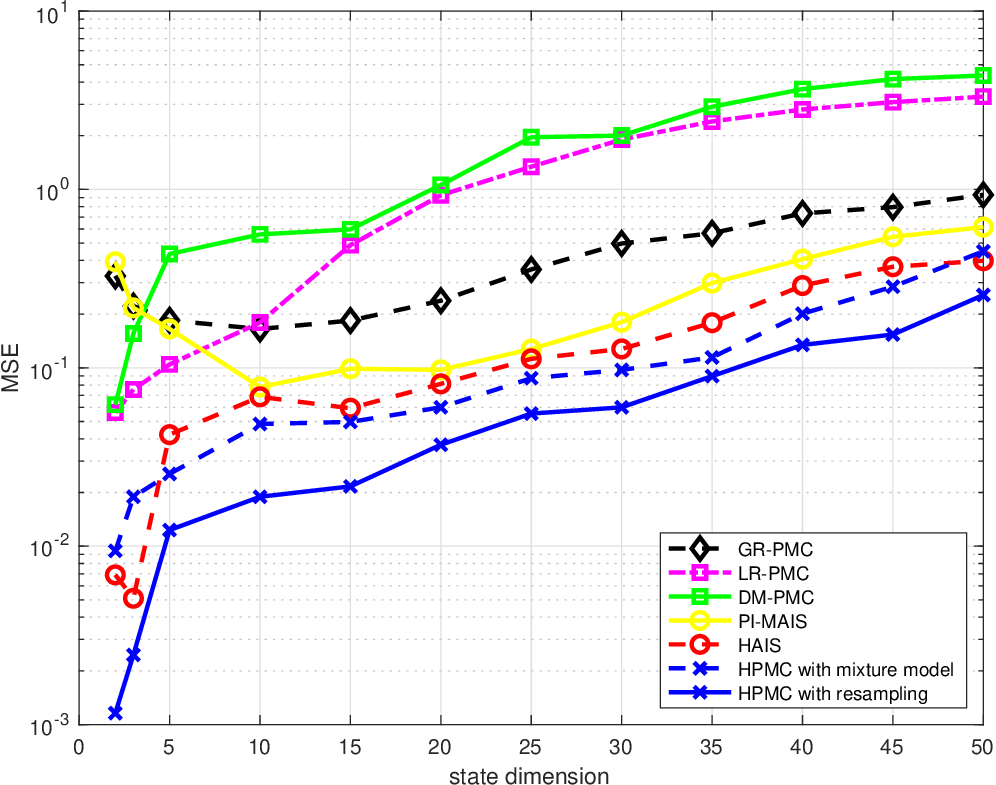}  
			\caption*{(a) $N=100$}
		\end{subfigure}
		\begin{subfigure}
			\centering
			\includegraphics[width=0.79\linewidth]{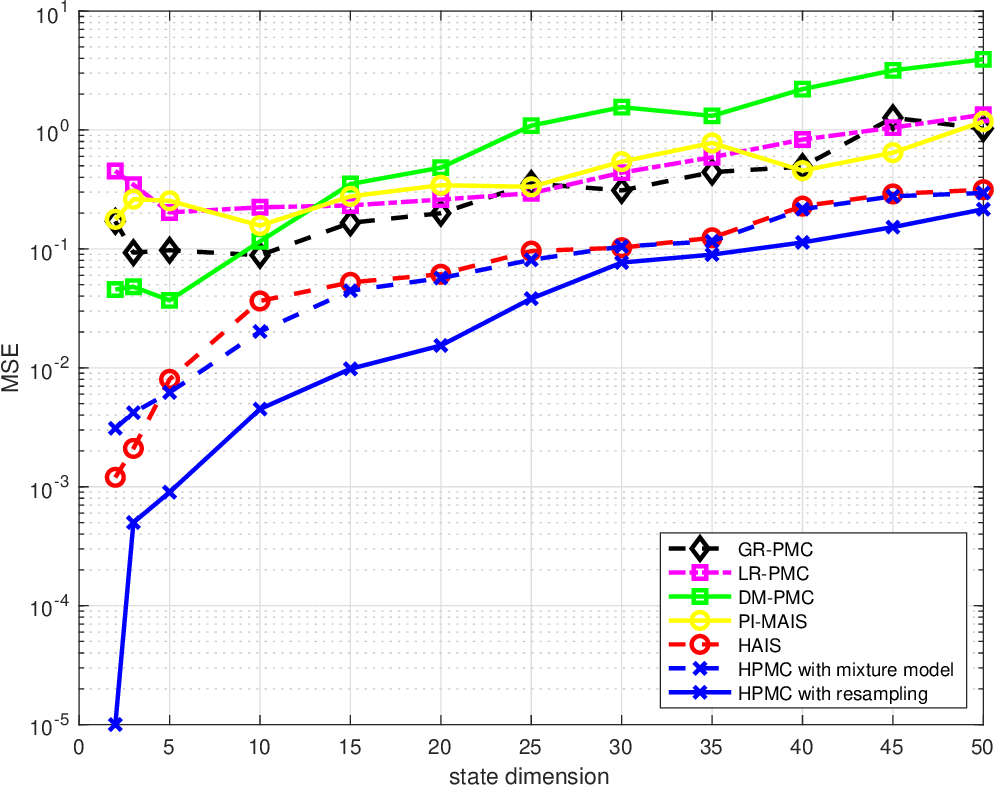}
			\caption*{(b) $N=200$}
		\end{subfigure}
		\caption{MSE of $ \mathbb{E}[\x] $, using two different values for $N$ as the number of proposals with $ \sigma=1 $, as the dimension of the target PDF, $ d_x $, increases.}
		\label{fig:Banana_Experiment}
	\end{figure}

	We now consider a benchmark multidimensional banana-shaped target distribution \cite{haario2001adaptive}, which is a challenging example because of its nonlinear nature, specially in high dimension. The target PDF is given by
	\begin{align}\nonumber
		&\bar{\pi}(x_1,...,x_{d_x}) \\
		&\propto \exp\left(-\frac{x_1^2}{2\sigma^2}-\frac{\left(x_2+b(x_1^2-\sigma^2)\right)^2}{2\sigma^2}-\sum_{i=3}^{d_x}\frac{x_i^2}{2\sigma^2}\right).
	\end{align}
	We set $ b=3 $ and $ \sigma=1 $, then the true value for $ \mathbb{E}_{\bar{\pi}}[\x]=0 $. In this experiment, we use Gaussian proposals with initial locations similar to those in the previous experiment. An isotropic covariance matrix is considered for all the proposal PDFs, $ \bm{\Sigma}=\sigma^2\bI_{d_x} $ with $ \sigma^2=1 $. 
	All algorithms are simulated with a fixed number of samples per proposals, i.e., $K=5$. However, we test two different values for the number of proposals, $N=100$ and $N=200$ to asses the impact of increasing $N$ on the performance of the estimator. 
	As in the previous experiment, to ensure a fair comparison, we select $T$ such that all algorithms have the same total number of target evaluations, $ E=2\times 10^5 $. We compute the MSE in the estimation of $ \mathbb{E}[\x] $ and the results are averaged over 200 Monte Carlo simulations. 
	To compare the performance of the proposed method with other approaches as the dimension of the state space increases, we vary the dimension of the state space by testing different values of $ d_x $ (with $ 2 \leq d_x \leq 50 $). 
	Fig. \ref{fig:Banana_Experiment} shows the MSE in the estimation of $ E[\x] $ as a function of the dimension $ d_x $ of the state-space, regarding the same techniques as in the previous bi-modal example. The performance of all methods degrades as the dimension increases.
	The result indicates that the novel HPMC with resampling scheme outperforms all the other methods under fair computational complexity comparison. In Fig. \ref{fig:Banana_Experiment}-b we doubled the number of proposal PDFs (to $N=200$) to see how the MSE values change across different dimensions. This adjustment leads to a significant decrease in MSE across all dimensions for all methods, with the proposed HPMC with the resampling method showing particularly notable improvements.
	
	\subsection{Computational complexity comparison }
	In this section, we provide a comparison of the computational complexity of different methods. In the MC domain, computational complexity is described as the total number of evaluations of proposal and target PDFs, as well as the computational cost per each generated sample.
	However, the number of target evaluations is of greater importance because, in most MC methods, the proposal PDF is typically chosen to be a Gaussian distribution, which is simpler to evaluate compared to complex target PDFs.
	Table \ref{table_Complexity} compares the computational complexity of different methods in terms of three parameters: $K$ (number of samples generated from each proposal PDF), $N$ (number of proposals PDFs), and $T$ (number of iterations). 
	In the DM weighting technique, $N$ proposal evaluations are required for each sample. As shown in Table \ref{table_Complexity}, all methods that use the DM weighting technique have $KN^2T$ as the number of proposal evaluations for the total $NKT$ generated samples.
	$3NT$ and $2NT$ are the number of target evaluations that occur in the cooperation by mixture model and resampling, respectively through iterations. As shown in the table, the number of target evaluations for the HPMC by resampling is lower than that for HPMC by mixture model. It can be noticed that the computational complexity of the proposed HPMC (mixture model) method is higher than that of all other algorithms.
	The other HPMC (with resampling), however, requires fewer computations, and its complexity is comparable to that of PI-MAIS and HAIS, two recent works in the literature.
	
	\begin{table*}[t!]
		\caption{Computational complexity comparison of different methods.}
		\label{table_Complexity}
		\centering
		\begin{tabular}{c c c c c}
			\hline 	\hline
			\multirow{3}{*}{Algorithm} & Number of & Number of & Number of &Number of \\
			& target evaluations & proposal evaluations & target evaluations & proposal evaluations \\
			& & & per sample & per sample\\
			\hline\hline
			Standard PMC & $NT$ & $NT$ & 1 & 1   \\
			DM-PMC  & $KNT$ & $KN^2T$ & 1 & $N$   \\
			N-PMC & $NT$ & $NT$ & 1 & 1  \\
			M-PMC & $KT$ & $KNT$ & 1 & $N$   \\
			LR-PMC & $KNT$ & $KN^2T$ & 1 & $N$   \\
			GR-PMC & $KNT$ & $KN^2T$ & 1 & $N$  \\
			AMIS & $KT$ & $KT^2$ & 1 & $T$ \\
			PI-MAIS & $KNT+NT$& $KN^2T$ & $1+1/K$ & $N$ \\
			APIS & $KNT$ & $KN^2T$ & 1 & $N$ \\
			GAPIS & $KNT$ & $KN^2T$ & 1 & $N$ \\
			HAIS & $KNT+NT$& $KN^2T$ & $1+1/K$ & $N$ \\
			HPMC (mixture model)& $KNT+3NT$&$KN^2T$&$1+2/K+1/KT$&$N$\\
			HPMC (resampling)& $KNT+2NT$&$KN^2T$&$1+2/K$&$N$\\
			\hline
		\end{tabular}
	\end{table*}
	
	\section{Conclusion}
	\label{sec_Conclusion}	
	In this paper, we have proposed the hybrid population Monte Carlo (HPMC) algorithm, a hybrid adaptive importance sampler with strong performance for high-dimensional and multi-modal problems. HPMC belongs to the population Monte Carlo (PMC) algorithms and incorporates a two-step hybrid adaptation mechanism. In the first step, preliminary locations are generated based on weighted samples and location parameter of the proposals. This hybrid approach enhances diversity, promotes both local and global exploration, and accelerates convergence. Notably, the novel cooperation step exchanges information among preliminary locations, further improving global exploration.
	The method is theoretically justified through its hybrid adaptation procedure and the cooperation techniques. Simulation results have shown significant improvements compared with other state-of-the-art methods. Future lines include the development of alternative adaptation mechanisms for the location parameters and the extension of the approach to adapt the covariance of the proposals.
	
	%\appendices
	%\label{appendix}
	%\bibliography{mybibs}{}

	\bibliographystyle{ieeetr}
	
\end{document}